\documentclass[usenatbib,a4paper]{mn2e}

\usepackage[psamsfonts]{amssymb}
\usepackage{graphics}

\def\tmag{{t_{\rm mag}}}

\def\rco{{R_{\rm co}}}

\def\mdot{{\dot{M}}}

\def\msun{{\rm M_{\odot}}}

\def\eg{{e.g.}}

\def\etal{{et al.}}

\def\Sig{{ \Sigma}}
\def\pie{{ \pi}}
\def\nue{{ \nu}}
\def\Lambd{{ \Lambda}}

\def\mstar{M_{\star}}

\def\rstar{R_{\star}}

\def\vr{v_{\rm {R}}}

\def\bphi{B_{{\varphi}}}
\def\bz{B_{{z}}}
\def\omegak{\Omega_{\rm{k}}}
\def\omegastar{\Omega_{\rm{\star}}}

\def\rc{r_{\rm{c}}}
\def\amag{a_{\rm{mag}}}
\def\cs{c_{\rm{s}}}
\def\rt{R_{\rm{t}}}

\begin{document}
\title[Steady-state accretion discs in central magnetic fields]
{The steady-state structure of accretion discs in central magnetic fields}
\author[O.M. Matthews et al.]{O.M. Matthews$^{1 \mbox{\footnotemark[1]\footnotemark[2]}}$, R. Speith$^2$, M.R. Truss$^3$ and G.A. Wynn$^1$ \\
$^1$ Department of Physics \& Astronomy, University of Leicester, University Road, Leicester, LE1 7RH \\
$^2$ Institut f{\"u}r Astronomie und Astrophysik, Universit{\"a}t T{\"u}bingen, Auf der Morgenstelle 10C, D-72076 T{\"u}bingen, Germany\\
$^3$ School of Physics \& Astronomy, University of St Andrews, North Haugh, St Andrews, Fife, KY16 9SS} 


\label{firstpage}

\maketitle

\begin{abstract}
We develop a new analytic solution for the steady-state structure of a
thin accretion disc under the influence of a magnetic field that is anchored
to the central star. The solution takes a form similar to that of
Shakura and Sunyaev and tends to their solution as the magnetic moment
of the star tends to zero. As well as the Kramer's law case, we obtain
a solution for a general opacity. The effects of 
varying the mass transfer rate, spin period and magnetic field of the star 
as well as the opacity model applied to the disc are explored for a range of objects. The solution depends on the position of the magnetic truncation radius. We propose a new approach for the identification of the truncation radius and present an analytic expression for its position.

\end{abstract}


\footnotetext[1]{E-mail: owen.matthews@astro.phys.ethz.ch}
\footnotetext[2]{Present address: Laboratory for Astrophysics, Paul Scherrer\\ Institut, W\"{u}renlingen und Villigen, CH-5232 Villigen PSI}

\section{Introduction}
\label{sec:int}

There is a well known one dimensional solution for the structure of a
thin accretion disc in the steady state. This solution, known as the
Shakura-Sunyaev disc solution \citep{sha73}, consists of a series of
seven equations. The full solution contains
expressions for disc surface density, scaleheight, density, temperature, 
opacity and viscosity. These expressions depend upon the Shakura-Sunyaev
alpha parameter, a dimensionless parametrization of viscosity. In addition to the alpha parameter the solution is
expressed as a function of radius, mass transfer
rate, stellar mass and stellar radius. 

We wish to reformulate 
the Shakura-Sunyaev disc solution to
incorporate the effect of a torque from a stellar magnetic field.  
A solution for the structure of an accretion disc in a magnetic field
is desirable because accretion discs are frequently found about
magnetic stars. Such magnetic fields can allow the transfer of angular
momentum between the disc and star and can therefore significantly
affect the structure of the disc as well as the spin evolution of the
star \citep[\eg][]{bra98,arm96}. These effects can be important in
young stars \citep*[\eg][]{kon90,arm99}, cataclysmic variables
\citep[\eg][]{sch02} and X-ray binaries \citep[\eg][]{pri72,rom04}. Accretion discs
can exhibit outbursts due to the thermal viscous disc instability, and the 
interoutburst time, known as the recurrence time, can be increased by 
several orders of magnitude by truncating the inner disc \citep*[\eg][]{mat03}. Accretion disc truncation by various mechanisms has also been invoked to explain the ultraviolet (UV) lag in dwarf novae \citep[\eg][]{mey94,kin97}. In this paper we treat magnetically enhanced accretion, but not the magnetic propeller effect.

A solution for the structure of a disc in a magnetic field is given by \citet{bra98}. By neglecting some mass transfer terms, they are able to find a detailed solution for the structure of the inner disc where a specific opacity model is assumed. In their solution magnetic pressure is considered in addition to magnetic torque, which improves the accuracy of the model for the inner disc regions. However, the azimuthal magnetic field is parametrized with respect to the shear between the disc and the stellar field using essentially the same form as is used in this paper. Since the effect of mass transfer into the disc is partially neglected, the solution is most valid in the very inner parts of the disc, where the magnetic field dominates the disc structure completely. A complementary solution is presented in this paper which, though less accurate in the centre of the disc, is continuous for the entire disc and can be applied to a range of opacity models. Furthermore the solution derived below tends to the Shakura-Sunyaev solution in the case where the stellar magnetic field vanishes, and also at large radii. 

The paper begins with a review of the necessary assumptions for the
new solution. Then an approximate formulation for the magnetic torque
is derived. This is applied, in conjunction with a new estimate of the magnetic truncation radius, to the derivation of a disc model. This model
is generalised in order that it is compatible with any opacity which
takes the form of a power law in density and temperature. We then
explore how variables such as the mass transfer rate, magnetic field and
stellar spin influence the disc structure.


\section{The solution}
\label{sec:str}

\subsection{Assumptions}
\label{sec:assump}

The derivation requires eight equations which follow in a form which
is appropriate for our new magnetic case. The equations used here are
similar to those presented by \citet*{fra01} in their derivation of the
standard Shakura-Sunyaev solution. First, density in the disc can be
approximated by 
\begin{equation}
\label{eqn:ssrho}
\rho = \frac{\Sigma}{H}
\;,
\end{equation}
where $\Sigma$ represents surface density and $H$ is the scaleheight
of the disc. For a thin disc the scaleheight can be approximated as  
\begin{equation}
\label{eqn:ssh}
H = \frac{\cs R^{3/2}}{\left(G \mstar \right)^{1/2}}
\;
\end{equation} 
where $\cs$ is the local sound speed, $R$ is the radial distance from the centre of the star in the plane of the disc and $\mstar$ is the mass of the primary star. The universal gravitational constant is represented by $G$. This approximation is likely to break down close to the star, where a strong magnetic field may cause gas to stream along magnetic field lines. An ideal gas has an equation of state of the form
\begin{equation}
\label{eqn:sscs}
\cs^{2} = \frac{P}{\rho}
\; ,
\end{equation}
in which $P$ represents pressure. Pressure is composed of three parts in the magnetic case: gas pressure, radiation pressure and magnetic pressure. We shall neglect the contributions of radiation and magnetic pressure, as these are much smaller than gas pressure for most cases. This yields
\begin{equation}\label{eqn:ssp}
P = \frac{\rho k T_{\rm c}}{\mu m_{\rm u}}
\; ,
\end{equation} 
where $k$ is the Boltzmann constant, $T_{\rm c}$ is the temperature in
the mid-plane of the disc, the atomic mass unit is given by 
$m_{\rm u}$, and $\mu$ is the mean molecular weight in the disc. This is
frequently taken to be $\mu \sim 0.6$. Conservation of energy gives
\citep{fra01}
\begin{equation}
\label{eqn:coe}
\frac{4 \sigma {T_{\rm c}}^4}{3 \tau} = \frac{9}{8} \nu \Sigma \frac{G \mstar}{R^{3}}
\; ,
\end{equation}
if it is assumed that the disc is Keplerian until the boundary layer and that the disc has a large optical depth. Here $\sigma$ represents the Stefan-Boltzmann constant and $\tau$ is the opacity of the disc material. At this stage the form of the function $\nu \Sigma$ is unknown, so that the equation must be kept in this general form. It is initially assumed that opacity may be approximated by Kramer's law such that
\begin{equation}
\label{eqn:kramer}
\tau = \kappa_{\rm{0}} \Sigma \rho T_{\rm c}^{-7/2} 
\quad\mbox{with}\quad
\kappa_{\rm{0}} = 
5 \times 10^{24} \mathrm{cm}^5\,\mathrm{g}^{-2}\,\mathrm{K}^{7/2}
\; . 
\end{equation}
Kramer's opacity is not a suitable model for all discs. Alternative opacity laws are discussed in Section {\ref{sec:opa}. Following the same approach as for the normal Shakura-Sunyaev solution, we assume that viscosity can be parametrized in the form  
\begin{equation}
\label{eqn:alphavisc}
\nu = \alpha c_{\rm s} H
\; .
\end{equation}
In addition, an expression for $\nu \Sigma$ must be obtained. In the
classical, non-magnetic Shakura-Sunyaev solution the relation 
\begin{equation}
\label{eqn:ssnusig}
\nu \Sigma = \frac{\mdot}{3 \pi}\left[1-\left(\frac{\rstar}{R} \right)^{1/2}\right]
\end{equation}
holds, where $\rstar$ denotes the radius of the star. However in the magnetic case the relation is more complex.

\subsection{Magnetic torque}
\label{sec:torque}

In order to obtain an expression for $\nu \Sigma$ in the magnetic case, we integrate the continuity equation in the direction normal to the plane of the disc, which gives
\begin{equation}\label{conteq}
\frac{\partial\Sig}{\partial t} + 
\frac{1}{R}\frac{\partial(R\Sig v_{\rm R})}{\partial R} = 0
\; .
\end{equation}
Similarly the radial component of the Navier-Stokes equation becomes
\begin{eqnarray}\label{radnavsto}
\lefteqn{\Sig\left(
\frac{\partial v_{\rm R}}{\partial t} + v_{\rm R}\frac{\partial v_{\rm R}}{\partial R}
- \frac{v_\varphi^2}{R}
\right)  = 
{}-\frac{\partial p}{\partial R} 
- \Sig\frac{G \mstar}{R^2}}
\\
\nonumber
 & &
{}+
\frac{4}{3R^{3/2}}\frac{\partial}{\partial R}
\left[R^{3/2}\nue\Sig\frac{\partial v_{\rm R}}{\partial R}\right]
-
\frac{2}{3R^3}\frac{\partial(R^2\nue\Sig v_{\rm R})}{\partial R}
\; ,
\nonumber
\end{eqnarray}
and the azimuthal component becomes
\begin{equation}\label{azinavsto}
\Sig\left(
\frac{\partial v_\varphi}{\partial t} + 
\frac{v_{\rm R}}{R}\frac{\partial(R v_\varphi)}{\partial R}
\right) =
\frac{1}{R^2}\frac{\partial}{\partial R}\left[
R^3\nue\Sig
\frac{\partial}{\partial R}\left(\frac{v_\varphi}{R}\right)\right]
\; ,
\end{equation}
where $v_{\rm \varphi}$ and $v_{\rm R}$ represent the azimuthal and radial
components of the velocity, respectively, and $\nue$ is the viscosity in the disc.

If we multiply equation (\ref{azinavsto}) by $R$, then we obtain an equation for the specific angular momentum,
\begin{equation}\label{angnavsto}
\Sig\left(
\frac{\partial l}{\partial t} + 
v_{\rm R}\frac{\partial l}{\partial R}
\right) =
\frac{1}{R}\frac{\partial}{\partial R}\left[
R^3\nue\Sig
\frac{\partial}{\partial R}\left(\frac{l}{R^2}\right)\right]
+ \Sig\Lambd
\; ,
\end{equation}
where we have added an external torque acting on
the disc, with $\Lambd$ representing the injection rate of specific angular momentum and $l$ indicating the current specific angular momentum. 
This follows the same procedure as \cite{lin86}, while a similar
method is also employed by \cite{pri91}.

Assuming that the disc is sufficiently cold and therefore that the 
dynamical time-scale is much smaller than the viscous time-scale, we can adopt a Keplerian approximation for the azimuthal motion such that $v_\varphi = (G \mstar /R)^{1/2}$. Equation
(\ref{angnavsto}) can then be solved for the radial velocity in the disc,
\begin{equation}\label{radvel}
v_{\rm R} =
-\frac{3}{R^{1/2}\Sig}
\frac{\partial(R^{1/2}\nue\Sig)}{\partial R}
+ 2\Lambd\frac{R^{1/2}}{\sqrt{G \mstar}}
\; .
\end{equation}
Inserting equation (\ref{radvel}) into the continuity equation
(\ref{conteq}) yields an evolution equation for the surface
density of the disc,
\begin{equation}\label{sigmaevol}
\frac{\partial\Sig}{\partial t} =
\frac{3}{R}\frac{\partial}{\partial R}\left[
R^{1/2}
\frac{\partial(R^{1/2}\nue\Sig)}{\partial R}
\right]
- 
\frac{1}{R}\frac{\partial}{\partial R}\left[
2\Lambd\Sig\frac{R^{3/2}}{\sqrt{G \mstar}}
\right]
\; ,
\end{equation}
where the right hand side is composed of a diffusion term, and an advection term that is due to the external torque.

To parametrize the specific torque $\Lambd$, we can write
\begin{equation}\label{torque}
\Lambd = \frac{l}{t_{\Lambd}} = \frac{\sqrt{G \mstar}R^{1/2}}{t_\Lambd}
\; ,
\end{equation}
where $t_\Lambd$ is the time-scale on which the local disc material
gains angular momentum.

In the present case, the source of the torque is the magnetic
interaction of a rotating, magnetic star with a partially ionised
disc. The torque time-scale is therefore equivalent to the magnetic
time-scale, $t_\Lambd\sim t_\mathrm{mag}$. 

The inclusion of
a magnetic field in such a hydrodynamic system introduces two additional
terms to the Euler equation: a magnetic pressure term and a magnetic 
tension term, \eg\ \citet{den90}. The magnetic pressure term is
negligible where $B$ is small. The magnetic tension term can be
expressed as 
\begin{equation}\label{btscl1}
\amag \sim \frac{1}{\rho \rc} \left( \frac{\bz \bphi}{4 \pie} \right)
\; ,
\end{equation} 
where $\rc$ represents the local
radius of curvature of the field lines as a result of the torque, while $\bz$ and $\bphi$ represent
the vertical and azimuthal components of the magnetic field
respectively. We use the approximation $\rc \sim H$
\citep*{pea97}. The ratio of vertical and azimuthal field strengths is
related to the shear between the disc and the magnetic field. If it is
assumed that the field rotates with the star then this ratio can be
expressed in the form \citep[\eg][]{liv92} 
\begin{equation}\label{btscl2}
\frac{\bphi}{\bz} \sim - \frac{\left(\omegak - \omegastar \right)}{\omegak}
\; ,
\end{equation}
where $\Omega_{\rm{k}}$ represents the angular frequency of a body in a Keplerian orbit at a given radius and $\Omega_{\rm{\star}} = \Omega_{\rm{B}}$ denotes the angular frequency of the star and therefore of the magnetic field at all radii. The magnetic time-scale can now be defined in terms of magnetic
acceleration and the Keplerian velocity by the relation 
\begin{equation}
\tmag \sim \frac{R \omegak}{\amag} \sim - \frac{4 \pie R^{2} \rho H}{
  \bz^{2}} \frac{\omegak^{2}}{\left({\omegak-\omegastar}\right)}
\; . 
\end{equation}
Finally, the volume density can be related to the surface density using equation (\ref{eqn:ssrho}) and for a
dipole field, we have $B_z \sim \left|\mbox{\boldmath{$\mu$}}\right|R^{-3}$ where $\mbox{\boldmath{${\mu}$}}$ is the magnetic moment of the
star. These relations lead to the following expression for the magnetic time-scale in the disc:
\begin{equation}
\label{tmagspec}
\tmag \sim \frac{4 \pie \Sig \sqrt{G \mstar} R^{11/2}}{\mbox{\boldmath{${\mu}$}}^{2} \left( \left[\frac{R}{\rco} \right]^{3/2} - 1 \right)}
\; ,
\end{equation}   
where $\rco$ represents the corotation radius, at which a body in a Keplerian orbit will revolve about the star at the same angular
frequency as that with which the star spins. If the field is moving more
rapidly than the disc, which occurs outside the corotation radius,
then the disc gains angular momentum and is pushed out to greater
radii. Inside the corotation radius the disc moves more rapidly than
the field and hence loses angular momentum and is accreted more
rapidly. These regions are known as magnetic propeller and magnetic
accretion regimes respectively. Because the magnetic propeller, at least in this one-dimensional treatment, completely suppresses accretion, the disc cannot reach a steady state in the propeller case. Our steady-state solution must therefore be confined to the magnetic accretion case. The result in equation
(\ref{tmagspec}) is almost identical to that adopted by
\citet{liv92}. It is possible to use other prescriptions for the
magnetic time-scale, for example the diamagnetic case 
\citep*[\eg][]{wyn97}. \citet{cam98a,cam98b} discuss the effects
of different assumptions for diffusivity. In general however $\tmag$
can usually be parametrized in the form
\begin{equation}
\label{tmaggen}
\tmag \sim \frac{2 \Sigma}{\beta} \frac{R^{\left( \gamma+2 \right)}}{\left( \left[\frac{R}{\rco} \right]^{3/2} - 1 \right)}
\; ,
\end{equation}
where $\gamma$ and $\beta$ are parameters determined by the magnetic
interaction model. In the fully magnetised case represented by
equation (\ref{tmagspec}) we have $\gamma = 7/2$ and $\beta$ is
determined by the mass and magnetic moment of the star
according to 
\begin{equation}
\label{betalast}
\beta \sim \frac{\mbox{\boldmath{${\mu}$}}^{2}}{2 \pi \sqrt{G \mstar}} 
\; .
\end{equation}
In fact the precise value of $\beta$ may vary by a factor of order unity. This is because the choice of a fully magnetised disc as well as the adoption of the relation $r_{c} \sim H$ are both simplifying assumptions.  

By considering equation (\ref{tmaggen}), a general expression for the
advection term in equation (\ref{sigmaevol}) is obtained which is similar
to that found by \citet{liv92}
\begin{equation}\label{genadvec}
\frac{1}{R}\frac{\partial}{\partial R}\left[
2\Lambd\Sig\frac{R^{3/2}}{\sqrt{G \mstar}}
\right]
=
\frac{{\it \beta}}{R}\frac{\partial}{\partial R}\left[
\frac{1}{R^{\gamma}}
\left(\left[\frac{R}{R_\mathrm{co}}\right]^{3/2} - 1\right)
\right]
\; .
\end{equation}
Equation (\ref{sigmaevol}) can now be solved analytically for the steady state.
By definition, in the steady state, the mass transfer rate $\mdot = - 2 \pi R \Sigma \vr$ is a constant
throughout the disc. By setting the left hand side of equation (\ref{sigmaevol}) to zero and integrating
the remaining terms, this condition can be used to derive a 
general solution for $\nu \Sigma$ in the steady state such that,
\begin{equation}\label{gensignew}
\nu \Sigma = \frac{\mdot}{3 \pi } +
  \frac{\beta R^{-\gamma}}{3 \left(2 - \gamma
    \right)}
\left(\left[\frac{R}{\rco} \right]^{3/2} - \frac{\left(2-\gamma \right)}{\left(\frac{1}{2} - \gamma \right)} \right)
+ C R^{- \frac{1}{2} }
\; ,
\end{equation}
where C is an arbitrary constant. 

\begin{figure*}
\begin{center}
\resizebox{88.1mm}{65.0mm}{
\mbox{
\includegraphics{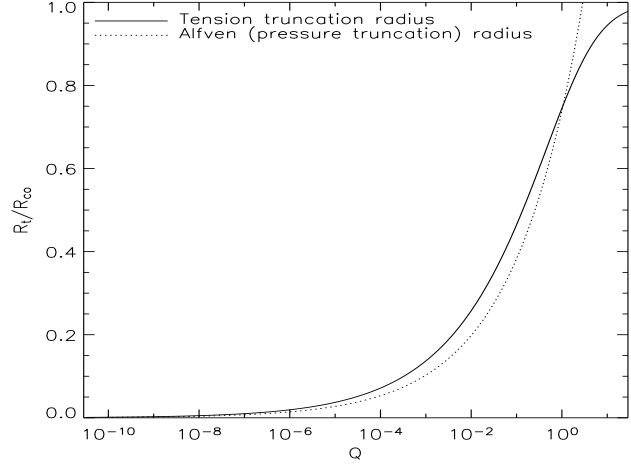}}}
\end{center}
\caption{Plot to show how the Alfv\'{e}n and tension truncation radii, as measured in units of corotation radii, vary as functions of the parameter Q, defined in equation (\ref{eqn:qdef}). The tension truncation radius is plotted for $\gamma=7/2$. Note that the tension truncation radius never exceeds corotation.}
\label{fig:trunc}
\end{figure*}

\subsection{Truncation radius}
\label{sec:truncation}
The truncation radius $R_{\rm{t}}$ is the radius within which the accretion rate is greatly enhanced by the magnetic field, and the surface density falls rapidly to a value close to zero. The position of this radius is of particular interest as it can provide a boundary condition for equation (\ref{gensignew}). Mathematically the truncation radius can be identified with the largest real root of equation (\ref{gensignew}). In the standard Shakura-Sunyaev solution the boundary condition $\nu \Sigma = 0$ is applied at the boundary layer, near $R=R_{\star}$, since there is no viscous transport at that point. Physically, in the magnetically truncated case, the surface density at radii smaller than the truncation radius must always be zero. Since equation (\ref{gensignew}) does not take this form, it is clear that the equation becomes unphysical where $R < R_{\rm{t}}$. It is therefore difficult to justify the use of the usual boundary condition in this case, since it will lie in an unphysical regime. If a truncation radius exists and can be identified, then it follows that $\nu \Sigma = 0$ at $R=R_{\rm{t}}$ may be used as a boundary condition.

There have been several attempts to locate the truncation radius due to the stellar magnetic field. The simplest of these estimates is the Alfv\'{e}n radius. This is the radius at which magnetic pressure is equal to ram pressure. According to \citet{fra01} the Alfv\'{e}n radius can be expressed by
\begin{equation}
\label{eqn:alfven}
R_{\rm{Alf}}=5.1 \times 10^{8} \mdot_{16}^{-2/7} {M_{1}}^{-1/7} {\mu_{30}}^{4/7} \ {\rm{cm}}
\; ,
\end{equation}
where $\mdot_{16}$ represents the mass transfer rate in units of $10^{16} \ {\rm{g \ s^{-1}}}$, $M_{1}$ is the stellar mass in solar masses and $\mu_{30}$ is the stellar magnetic moment in units of $10^{30} {\rm{G \ cm^{3}}}$. This equation is not sensitive to the spin of the star. Indeed by inspection we see that the Alfv\'{e}n radius can be related to $\beta / \mdot$ so that
\begin{equation}
R_{\rm{Alf}}=1.25 \left( \beta / \mdot \right)^{2/7}\mathrm{cm}
\; .
\label{eqn:alfven2}
\end{equation}
A more sophisticated approach than that of the Alfv\'{e}n radius is adopted by \citet{bra98}, who include magnetic tension and magnetic pressure in their calculations. Since some mass transfer terms are neglected, their solution is most valid very close to the star. \citet{bra98} do not provide an analytic formulation for the truncation radius, but according to their data table the radii that they calculate are all close to the Alfv\'{e}n radius. 

The approach adopted here, which is consistent with section \ref{sec:torque} in that it treats the magnetic tension only, is to solve equation (\ref{gensignew}) for $R_{\rm{t}}$ by taking advantage of the fact that the inner accretion disc is empty in the case of magnetically enhanced accretion. If $\Sigma \left( R_{\rm{t}} \right) = 0$, and if annuli at smaller radii also have a zero surface density then, in addition, $\nu \Sigma = 0$ at $R = R_{\rm{t}}$ and at immediately smaller radii. It is reasonable to conclude from this that $\partial \left( \nu \Sigma \right)/ \partial R = 0$ at $R=R_{\rm{t}}$. Physically this can be interpreted as the effect of a very rapid magnetic accretion. The inner edge of the disc loses angular momentum on a short time-scale and is dragged towards the stellar surface so rapidly that the inner disc has a negligible surface density. These two conditions for $\nu \Sigma$ are sufficient to obtain a relationship between mass transfer rate, $\beta$, truncation radius and corotation radius from equation (\ref{gensignew}) so that
\begin{equation}
\label{eqn:newtrunc1}
\label{eqn:qdef}
\frac{\mdot}{\beta} = 2 \pi R_{\rm{t}}^{- \gamma} \left( 1 - \left( \frac{R_{\rm{t}}}{R_{\rm{co}}} \right)^{3/2} \right) 
\; .
\end{equation}
This result can already tell us something about the behaviour of the truncation radius in this approach. It is clear that when $\mdot / \beta = 0$, and the magnetic effect is at its most extreme, we have $R_{\rm{t}}=R_{\rm{co}}$. It is not possible for $R_{\rm{t}}$ to exceed $R_{\rm{co}}$ providing that  $\mdot / \beta \geq 0$. This agrees with our previous assertion that the magnetic propeller does not have a steady-state solution. Of course when $\beta / \mdot = 0$ there is no truncation and we find that $R_{\rm{t}} = 0$. In reality however, since the inner radius of the disc cannot be smaller than the stellar radius $\rstar$, the minimum value for the truncation radius is always $R_{\rm{t}}=\rstar$. It is also clear that, according to this formulation, $R_{\rm{t}}$ is independent of the disc opacity, and of $\alpha$ except in as much as $\mdot$ is related to $\alpha$. There is no general analytic solution for $R_{\rm{t}}$ from equation (\ref{eqn:newtrunc1}), however it is possible to obtain a solution for certain values of $\gamma$. In order to do this, and to interpret the result, it is useful to make a parametrization of equation (\ref{eqn:newtrunc1}). We define 
\begin{equation}
\label{eqn:newtrunc2}
Q=2 \pi R_{\rm{co}}^{-\gamma} \left( \frac{\beta}{\mdot} \right)
\; .
\end{equation}
Combining this with equation (\ref{eqn:newtrunc1}) yields
\begin{equation}
\label{eqn:newtrunc3}
Q = \frac{\left( \frac{R_{\rm{t}}}{R_{\rm{co}}} \right)^{\gamma}}{1- \left( \frac{R_{\rm{t}}}{R_{\rm{co}}} \right)^{3/2}}
\; .
\end{equation}
The tension truncation radius given by equation (\ref{eqn:newtrunc3}), and with $\gamma=7/2$, is plotted in Fig. \ref{fig:trunc}. The Alfv\'{e}n radius, which is also plotted, is smaller than the tension truncation radius for
much of the regime of interest. Where the Alfv\'{e}n radius is
greater than the tension truncation radius the magnetic pressure may
be said to dominate. However, when the Alf\'{e}n radius exceeds the corotation radius, it cannot represent the truncation radius of a steady-state accretion disc. For this reason the truncation radius given by equation (\ref{eqn:newtrunc3}) is preferable.

The great advantage of the $Q$ parametrization is that once the value of Q has been established from equation (\ref{eqn:newtrunc2}) it is straightforward to read off the value of $R_{\rm{t}}/R_{\rm{co}}$, which can only vary from zero to unity. This can be done either from Fig. \ref{fig:trunc} or from similar plots for different values of $\gamma$. This means that only one plot is required for all the parameter space of interest. For the fully magnetised case, where $\gamma = 7/2$, an exact analytic solution cannot be found for $R_{\rm{t}}/R_{\rm{co}}$ from equation (\ref{eqn:newtrunc3}). In the region where $R_{\rm{t}}/R_{\rm{co}}$ is close to unity, the $\gamma=7/2$ case can be closely approximated by setting $\gamma=3$, which yields a quadratic equation in $\left(R_{\rm{t}}/R_{\rm{co}}\right)^{3/2}$. This, of course, has an exact analytic solution, and since this is very close to the solution for $\gamma=7/2$ for all of the region of interest, we can make for that case the approximation
\begin{equation}
\label{eqn:newtrunc4}
\frac{R_{\rm{t}}}{R_{\rm{co}}} \simeq \left(\frac{Q}{2}\left(\sqrt{\frac{4}{Q}+1}-1\right)\right)^{2/3}
\; .
\end{equation}
While the fractional error in this approximation becomes larger when $R_{\rm{t}}/R_{\rm{co}}$ approaches zero, the absolute error is always less than $0.05$, and, in any case $R_{\rm{t}}/R_{\rm{co}}$ can never become very small as physically it should be limited by $\rstar$.

} 


\subsection{Synthesis}
\label{sec:synthesis}

The result in equation (\ref{gensignew}) was obtained in
\citet{mat03}. However in that work the standard non-magnetic
Shakura-Sunyaev viscosity prescription was substituted into the
equivalent of equation (\ref{gensignew}) in order to obtain a surface
density profile. This is clearly not self-consistent. In this paper no
assumption of the form of $\nu$ will be made other than that in
equation (\ref{eqn:alphavisc}). \citet{mat03} also used an arbitrary inner
boundary condition such that $\partial \Sigma / \partial R = 0$ at $R=R_{\star}$. In the
following derivation the inner boundary condition $\Sigma \left( R_{\rm{t}} \right) = 0$, which is equivalent to the condition $\nu \Sigma = 0$ at $R=R_{\rm{t}}$, will be
used, as discussed in Section \ref{sec:truncation}. The value of $R_{\rm{t}}$ can be obtained using equation (\ref{eqn:newtrunc4}) or by another method. In any case, where $R_{\rm{t}} < R_{\star}$, the boundary condition $\Sigma \left( R_{\star}\
 \right) = 0$ should be used instead, in agreement with the usual approach
\citep[\eg][]{fra01}. The above boundary
condition for $R_{\rm{t}} > R_{\star}$ yields an expression for the constant C, when applied to
equation (\ref{gensignew}), so that
\begin{equation}
\label{eqn:fullnusig}
\nu \Sigma = \frac{\mdot}{3
  \pi}\left(1-\left[\frac{R_{\rm{t}}}{R}\right]^{1/2}\right)-\frac{\beta
  R^{-\gamma}}{3\left(\gamma-2 \right)}h
= \frac{\mdot}{3\pi}f^4v
\; ,
\nonumber
\end{equation}
where 
\begin{eqnarray}
\label{eqn:hdef}
h & = & 
\left[\frac{R}{\rco}\right]^{\frac{3}{2}}\left(1-\left[{\frac{R_{\rm{t}}}{R}}\right]^{\left(2 - \gamma
  \right)}\right)
\nonumber\\
& &
{} -\frac{\left(\gamma-2\right)}{\left(\gamma-\frac{1}{2}\right)}\left(1-{\left[\frac{R_{\rm{t}}}{R}\right]}^{\left(\frac{1}{2}-\gamma\right)}\right)
\\
\label{eqn:fdef}
f & =& \left(1 - \left[ \frac{R_{\rm{t}}}{R} \right]^{1/2} \right)^{1/4}
\; ,
\\
\label{eqn:defv}
v 
& = & 
1 - \frac{\beta}{\mdot}\frac{\pi h}{R^{\gamma }\left(\gamma -2 \right) f^{4}} 
\; ,
\end{eqnarray}
where $R_{\star}$ replaces $R_{\rm{t}}$ throughout if $R_{\rm{t}} < R_{\star}$. In the limit $\beta = 0$, this therefore yields the non-magnetic result in equation (\ref{eqn:ssnusig}) as expected. 
Substituting relation (\ref{eqn:fullnusig}) into equation
(\ref{eqn:coe}) gives  
\begin{equation}
\label{eqn:coe2}
\frac{4 \sigma {T_{\rm c}}^4}{3 \tau} = 
\frac{9}{8} \frac{G \mstar}{R^{3}} \frac{\mdot}{3\pi}f^4v
\; .
\end{equation}
Following a procedure similar to that of \citet{fra01}, we can now
obtain a full analytic solution for the thin accretion disc, to which
a magnetic torque is applied. 
Solving the set of algebraic equations (\ref{eqn:ssrho})-(\ref{eqn:ssp}), (\ref{eqn:kramer}), (\ref{eqn:alphavisc}), 
(\ref{eqn:fullnusig}) and (\ref{eqn:coe2}) yields 
the following expression for surface density: 
\begin{eqnarray}\label{eqn:longsigma}
\Sigma & = &
\left(\frac{32\sigma}{27\kappa_0}\right)^{1/10}\left(\frac{\mu
m_{\rm{u}}}{k}\right)^{3/4} 
\nonumber\\
& &
\alpha^{-4/5} \left( {\frac{G
\mstar}{R^{3}}} \right)^{1/4} 
\left({\frac{\mdot f^{4}}{3 \pi} v} \right)^{7/10}
\; .
\end{eqnarray}
If $\mu = 0.6$ then the result can be rearranged into a more familiar form so that
\begin{equation}
\label{eqn:mtsig}
\Sigma = 3.7 \alpha^{-4/5} \mdot_{16}^{7/10} M_{1}^{1/4} R_{10}^{-3/4} f^{14/5} v^{7/10} {\rm{g \ cm^{-2}}}
\; .
\end{equation} 
The mass transfer rate in units of $10^{16} \ {\rm g \ s^{-1}}$ is
represented by $\mdot_{16}$ and $M_{1}$ is the mass of the accreting
star in solar masses. The radial distance from the star, in the plane
of the disc, is represented by $R$ in units of $10^{10} \ {\rm
cm}$. The numerical coefficient is sensitive to the values adopted for
both $\mu$ and for opacity. Other than this the result in equation
(\ref{eqn:mtsig}) differs from the usual form only by the presence of
the correction term $v$, and the substitution of $R_{\rm{t}}$ for $R_{\star}$ throughout. It is instructive that the correction term
differs from unity by an amount which is proportional to the ratio of
$\beta$ to the mass transfer rate. This means that, as would be
expected, a sufficiently high mass transfer rate can overcome the
magnetic field of a star and the disc will revert to a Shakura-Sunyaev
form. 

The remaining parts of the solution are also easily found and are
collected below. In all cases they differ from the usual
Shakura-Sunyaev form by the presence of the factor $v$: 
\begin{equation}
\label{eqn:mth}
H = 1.7 \times 10^{8} \alpha^{-1/10} \mdot_{16}^{3/20} M_{1}^{-3/8} R_{10}^{9/8} f^{3/5} v^{3/20} {\rm{cm}}
\end{equation} 
\begin{equation}
\label{eqn:rhomt}
\rho = 3.7 \times 10^{-8} \alpha^{-7/10} \mdot_{16}^{11/20} M_{1}^{5/8} R_{10}^{-15/8} f^{11/5} v^{11/20} {\rm{g \ cm^{-3}}}
\end{equation}
\begin{equation}
\label{eqn:tcmt}
T_{c} = 2.5 \times 10^{4} \alpha^{-1/5} \mdot_{16}^{3/10} M_{1}^{1/4} R_{10}^{-3/4} f^{6/5} v^{3/10} {\rm{K}}
\end{equation}
\begin{equation}
\label{eqn:taumt}
\tau = 180 \alpha^{-4/5} \mdot_{16}^{1/5} f^{4/5} v^{1/5} 
\end{equation}
\begin{equation}
\label{eqn:numt}
\nu = 2.9 \times 10^{14} \alpha^{4/5} \mdot_{16}^{3/10} M_{1}^{-1/4} R_{10}^{3/4} f^{6/5} v^{3/10} {\rm{cm^{2} \ s^{-1}}}
\end{equation}
The radial velocity of the disc material can be calculated 
directly from 
\begin{equation}
\label{eqn:vrmt}
v_{\rm{R}} = -\frac{\mdot}{2\pi}R^{-1}\Sigma^{-1}
\; .
\end{equation}

\subsection{The general opacity case}
\label{sec:var}
\label{sec:opa}

Kramer's opacity is valid for discs where $T_{\rm{c}} \gtrsim 1 \times 10^{4} \ {\rm{K}}$. For discs around young stars, for example, this condition will not always be valid. The range of opacities in such discs is discussed in \citet{sem03}. It is therefore useful to examine some alternative opacity laws and apply them in a manner similar to that found above. The process is straightforward, providing the opacity can be approximated by the form
\begin{equation}
\label{eqn:opacitygen}
\tau = \kappa_{\rm{0}} \Sigma \rho^{\rm{a}} {T_{\rm{c}}}^{\rm{b}}
\; ,
\end{equation}
where $\kappa_{\rm{0}}$, $a$ and $b$ are constants. A collection of
opacity prescriptions in this form can be found in the appendix of
\cite{bel94}. 
The equations of the full solution are collected below, where $d=20+6a-4b$. 
\begin{eqnarray}
\label{eqn:arbsig}
\Sigma = \left(\frac{32 \sigma}{27 \kappa_{0}}\right)^{\frac{4}{d}}
\left(\frac{\mu m_{\rm{u}}}{k}\right)^{\frac{16-4b}{d}}
\left(\frac{G\mstar}{R^{3}}\right)^{\frac{4-a-2b}{d}} \nonumber \\
\alpha^{\frac{-16-2a+4b}{d}}
\left({\frac{\mdot f^{4}}{3 \pi} v} \right)^{\frac{12+2a-4b}{d}}
\end{eqnarray}

\begin{eqnarray}
\label{eqn:arbh}
H = \left(\frac{32 \sigma}{27\kappa_{0}}\right)^{\frac{-2}{d}}
\left(\frac{\mu m_{\rm{u}}}{k}\right)^{\frac{-8+2b}{d}}
\left(\frac{G\mstar}{R^{3}}\right)^{\frac{-7-a+2b}{d}} \nonumber \\
\alpha^{\frac{-2-2a}{d}}
\left({\frac{\mdot f^{4}}{3 \pi} v} \right)^{\frac{4+2a}{d}}
\end{eqnarray}

\begin{eqnarray}
\label{eqn:arbrho}
\rho = \left(\frac{32\sigma}{27\kappa_{0}}\right)^{\frac{6}{d}}
\left(\frac{\mu m_{\rm{u}}}{k}\right)^{\frac{24-6b}{d}}
\left(\frac{G\mstar}{R^{3}}\right)^{\frac{11-4b}{d}} \nonumber \\
\alpha^{\frac{-14+4b}{d}}
\left({\frac{\mdot f^{4}}{3 \pi} v} \right)^{\frac{8-4b}{d}}
\end{eqnarray}

\begin{eqnarray}
\label{eqn:arbtc}
T_{\rm{c}} = \left(\frac{32\sigma}{27\kappa_{0}}\right)^{\frac{-4}{d}}
\left(\frac{\mu m_{\rm{u}}}{k}\right)^{\frac{4+6a}{d}}
\left(\frac{G\mstar}{R^{3}}\right)^{\frac{6+4a}{d}} \nonumber \\
\alpha^{\frac{-4-4a}{d}}
\left({\frac{\mdot f^{4}}{3 \pi} v} \right)^{\frac{8+4a}{d}}
\end{eqnarray}

\begin{eqnarray}
\label{eqn:arbtau}
\tau = \kappa_{0}\left(\frac{32\sigma}{27\kappa_{0}}\right)^{\frac{4+6a-4b}{d}}
\left(\frac{\mu m_{\rm{u}}}{k}\right)^{\frac{16+24a}{d}}
\left(\frac{G\mstar}{R^{3}}\right)^{\frac{4+10a+4b}{d}} \nonumber \\
\alpha^{\frac{-16-16a}{d}}
\left({\frac{\mdot f^{4}}{3 \pi} v} \right)^{\frac{12+10a+4b}{d}}
\end{eqnarray}

\begin{eqnarray}
\nu = \alpha\left(\frac{32\sigma}{27\kappa_{0}}\right)^{\frac{-4}{d}}
\left(\frac{\mu m_{\rm{u}}}{k}\right)^{\frac{-16+4b}{d}}
\left(\frac{G\mstar}{R^{3}}\right)^{\frac{-4+a+2b}{d}} \nonumber \\
\alpha^{\frac{16+2a-4b}{d}}
\left({\frac{\mdot f^{4}}{3 \pi} v} \right)^{\frac{8+4a}{d}}
\label{eqn:arbnu}
\end{eqnarray}

\section{Form of functions}
\label{sec:form}
\begin{figure*}
\begin{center}
\resizebox{88.1mm}{65.0mm}{
\mbox{
\includegraphics{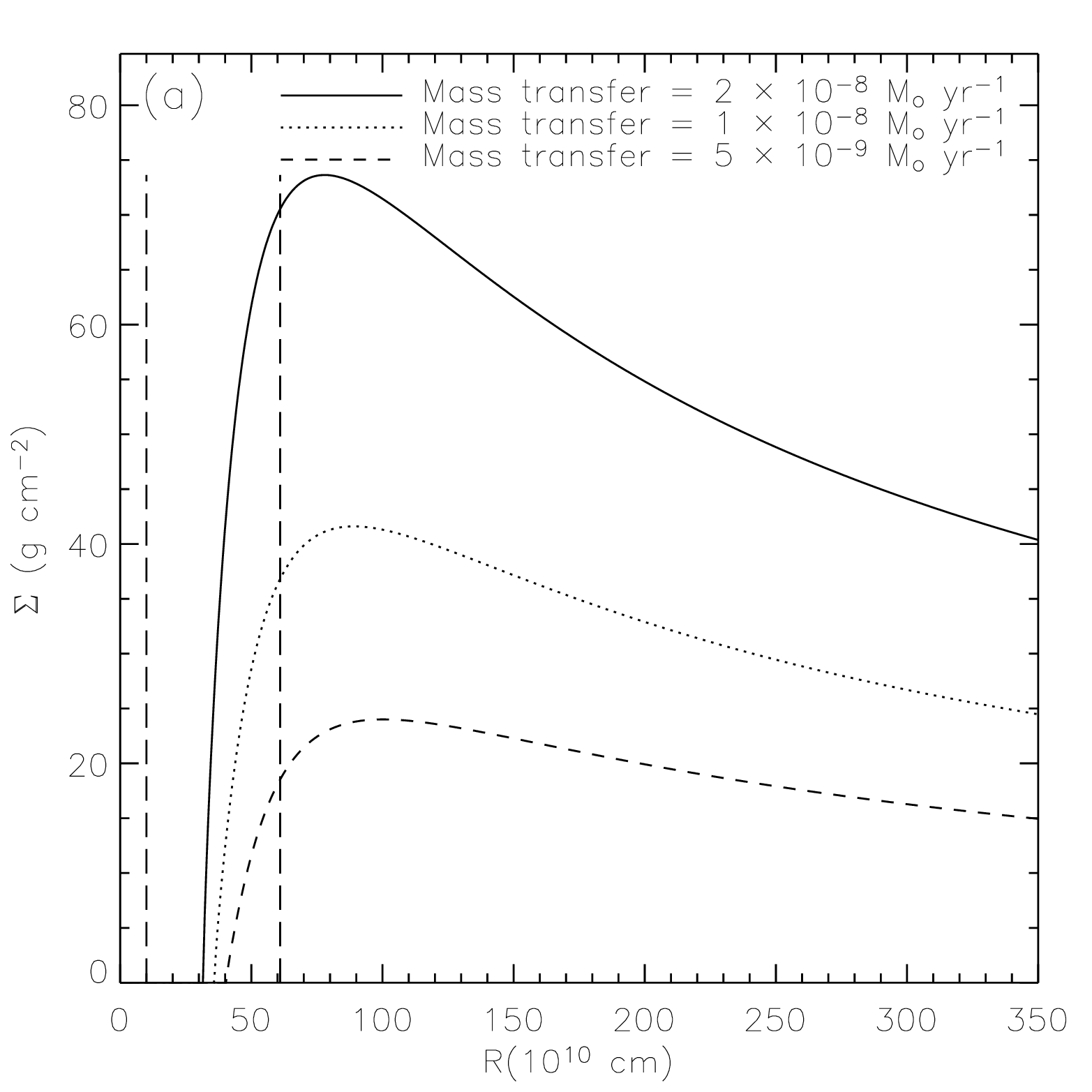}}} 
\resizebox{88.1mm}{65.0mm}{
\mbox{
\includegraphics{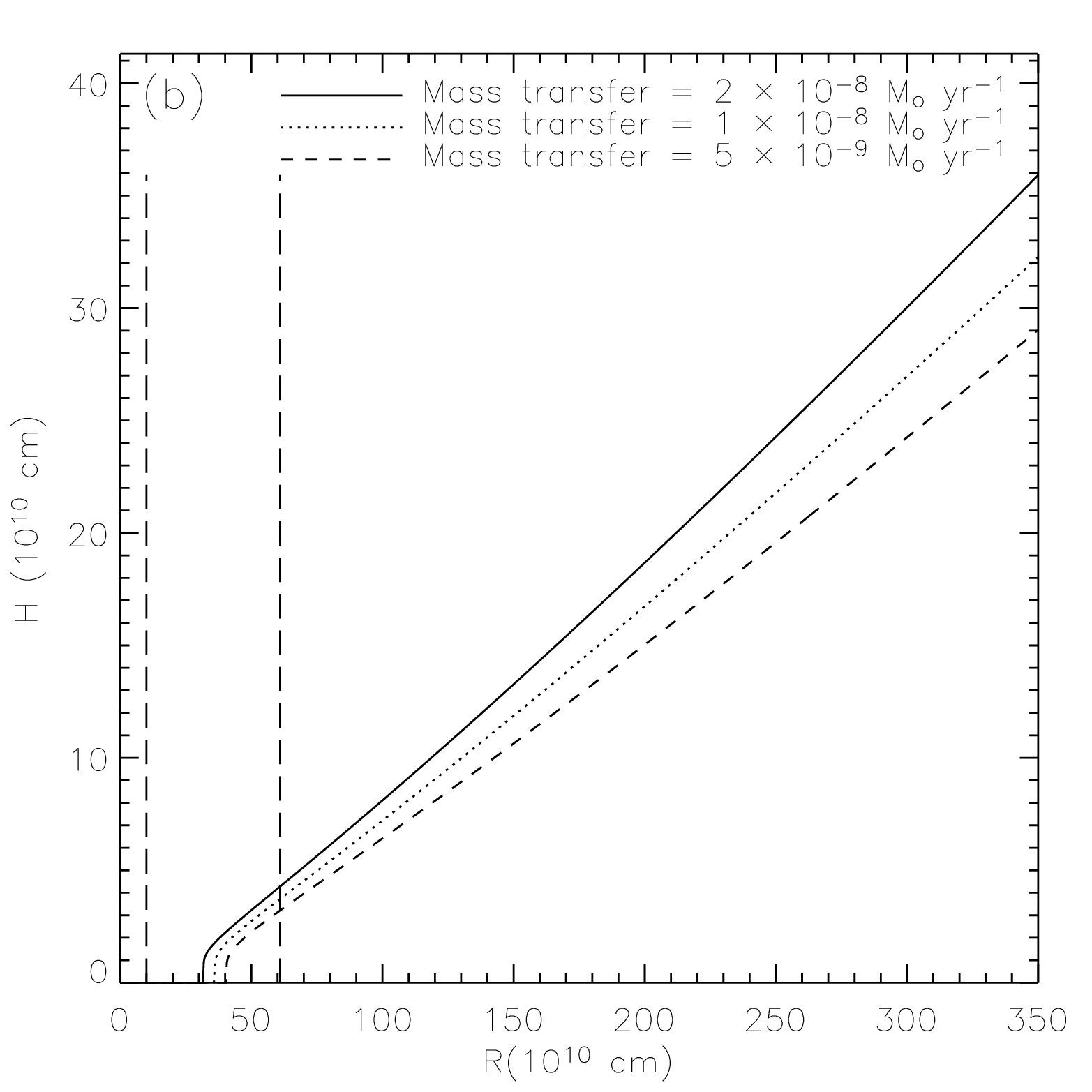}}} 
\resizebox{88.1mm}{65.0mm}{
\mbox{
\includegraphics{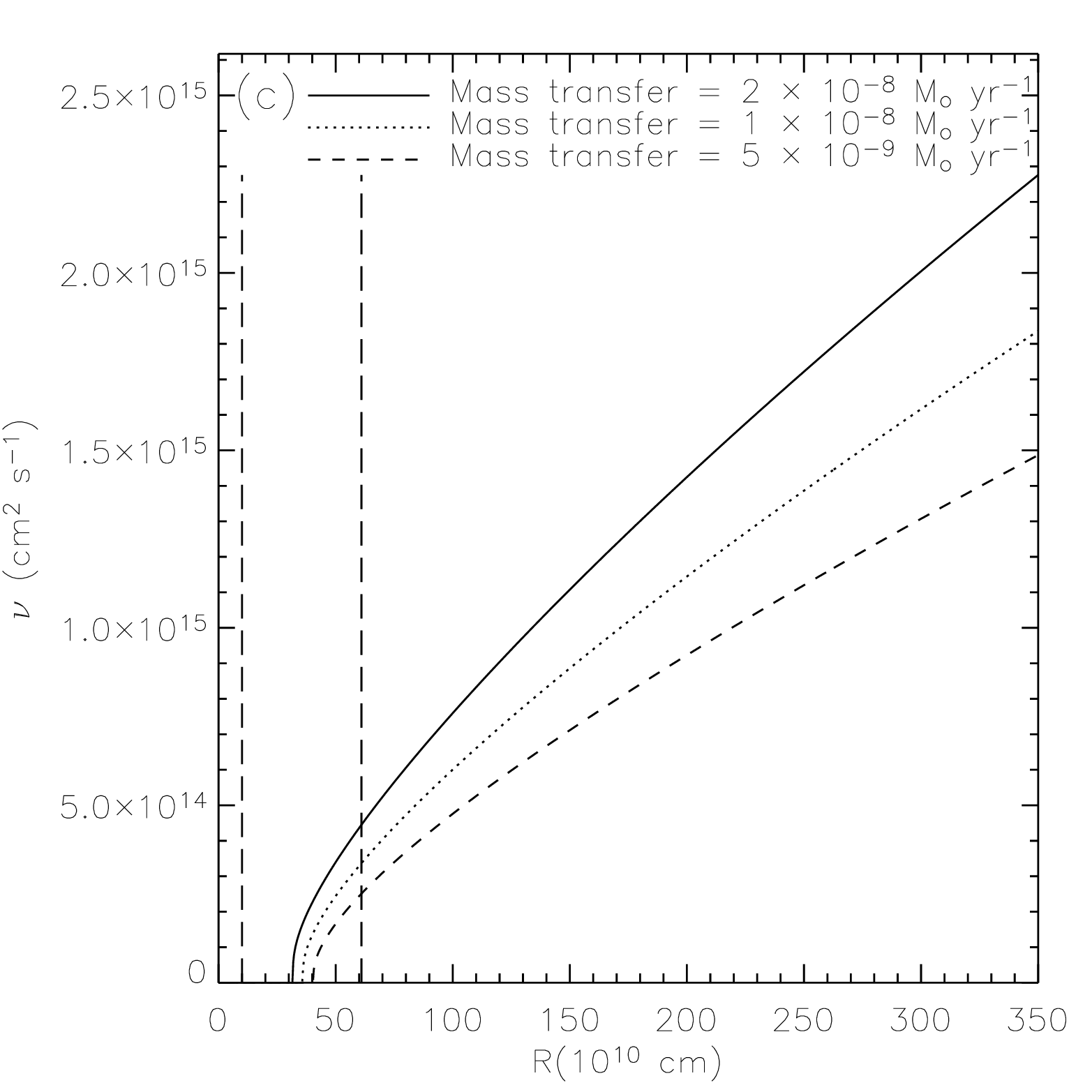}}}
\resizebox{88.1mm}{65.0mm}{
\mbox{
\includegraphics{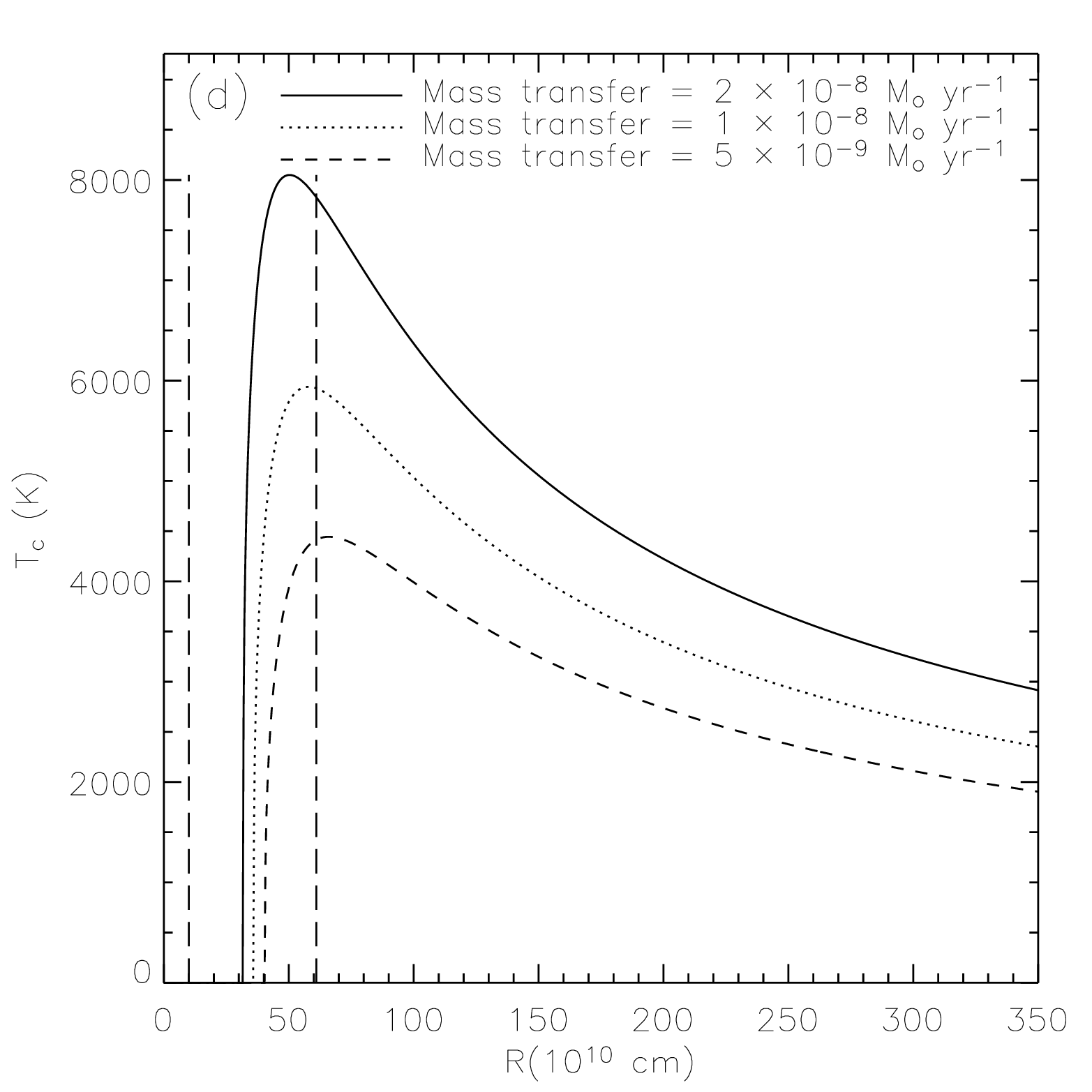}}}
\resizebox{88.1mm}{65.0mm}{
\mbox{
\includegraphics{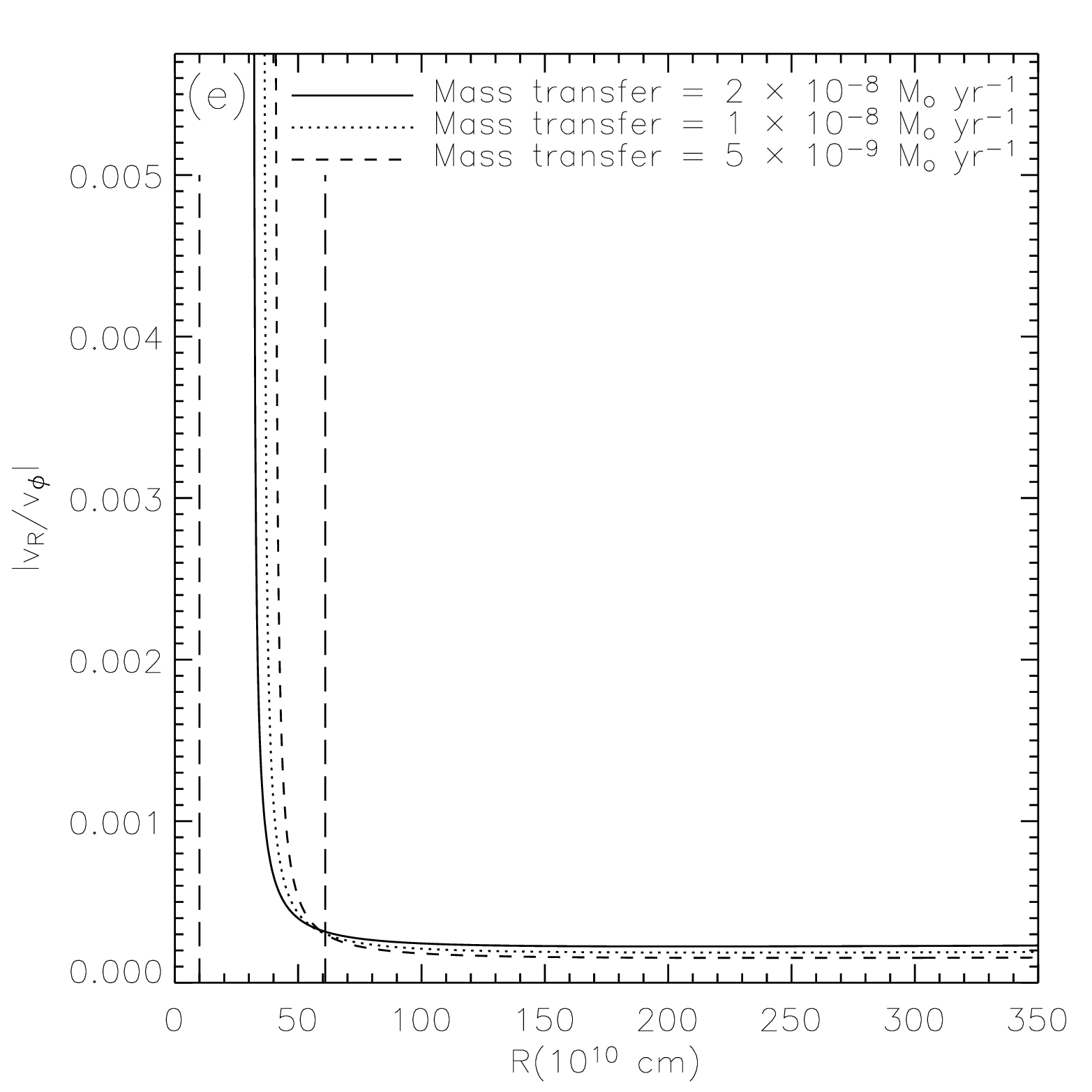}}}
\end{center}
\caption{Analytic model of a typical YSO disc showing the surface
density, disc scaleheight, viscosity and central temperature as a
function of radius. The ratio of radial to azimuthal velocity is 
also shown. The disc is plotted from $R=0$ to $R=3.5 \times
10^{12} \ {\rm{cm}}$. The star has a radius $\rstar=1 \times 10^{11} \
{\rm{cm}}$, mass $\mstar=1 \ \msun$, spin period $P_{\rm{spin}}=3 \
{\rm{d}}$ and a surface magnetic field of $B=500 \ {\rm{G}}$. The mass transfer rate throughout the disc are from $\mdot=5 \times 10^{-9} \ \msun \
{\rm{yr}}^{-1}$ to $2 \times 10^{-8} \ \msun \ {\rm{yr}}^{-1}$.  An alpha parameter of $\alpha = 0.01$ is adopted, and Kramer's opacity is assumed. The vertical lines illustrate the stellar radius and the corotation radius.}
\label{fig:one}
\end{figure*}

Some typical disc structures will now be illustrated, using the full
disc solution quoted in equations (\ref{eqn:arbsig})-
(\ref{eqn:arbnu}). The effect upon the solution of altering some of
the parameters will also be shown. In every case the truncation radius, which is required for the solution, is obtained from equation (\ref{eqn:newtrunc4}). It is therefore implicitly assumed that $\gamma=7/2$. 
Fig. \ref{fig:one} shows the
solution as applied to a disc surrounding a typical young stellar
object (YSO). The disc is plotted from $R=0$ to $R=3.5 \times 10^{12} \
{\rm{cm}}$. This is an arbitrary maximum radius, used to illustrate
the behaviour of the inner disc since, in a real YSO, the disc would be much larger than this. In the case of a single YSO, which occupies a simple potential,
the model can be extended outwards indefinitely. It should be
noted however that, in a model with a great radial extent, it is
unlikely that the entire disc could be correctly represented with a
single $\alpha$ parameter \citep[\eg][]{gam96}, and a single opacity
prescription. A further complication in these massive, relatively cool
discs, is that the gravitational instability may also contribute
towards the disc viscosity. In the present case however, where only
the inner disc is modelled, Kramer's opacity is assumed throughout the
simulated region and a global value of $\alpha = 0.01$ is
adopted. This is undoubtedly a simplification, but is adequate for the
purpose of illustration. This value of $\alpha$ in YSOs is
consistent with FU Orionis recurrence times in the magnetic case. 
\citep{mat03}. 

\begin{figure*}
\begin{center}
\resizebox{88.1mm}{65.0mm}{
\mbox{
\includegraphics{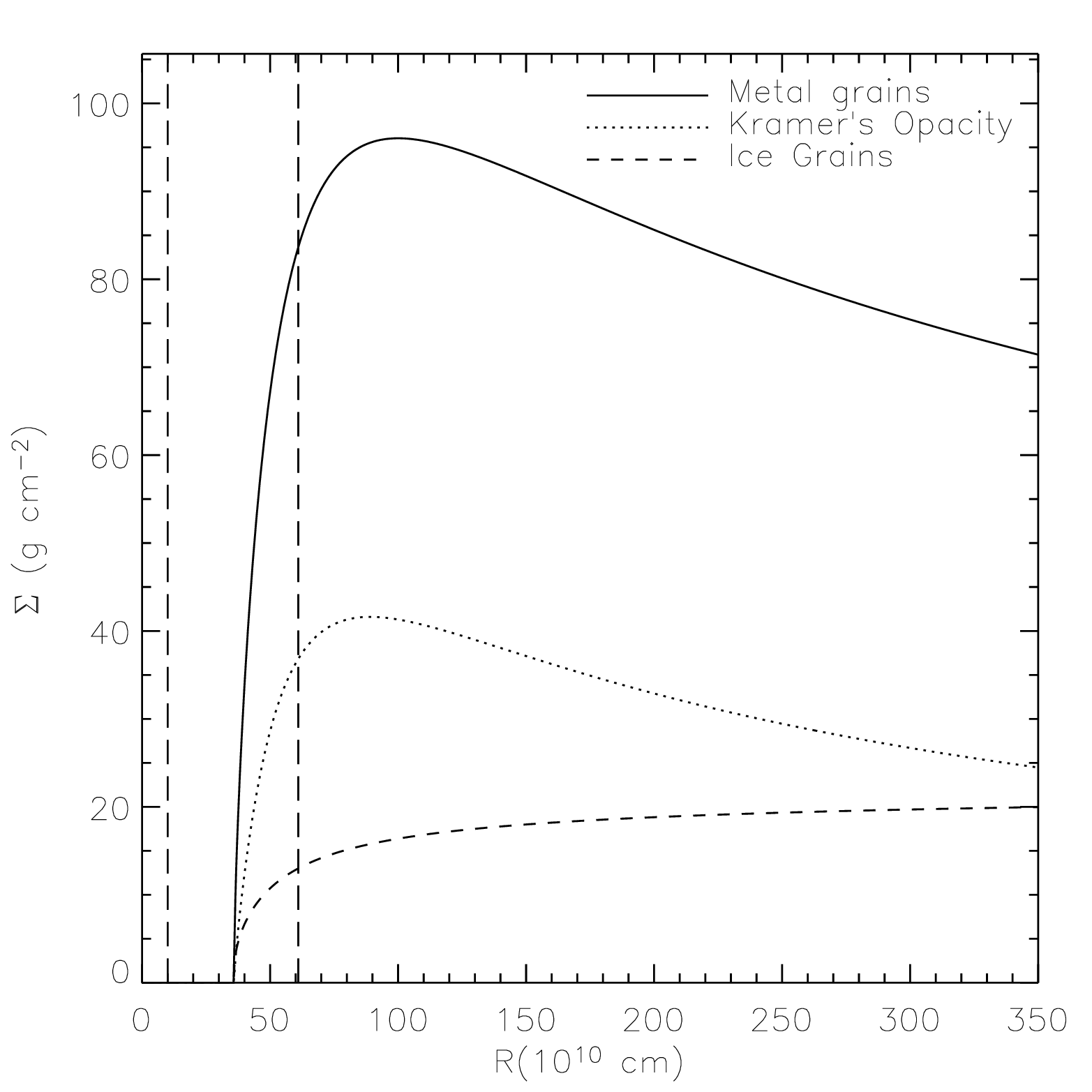}}} 
\resizebox{88.1mm}{65.0mm}{
\mbox{
\includegraphics{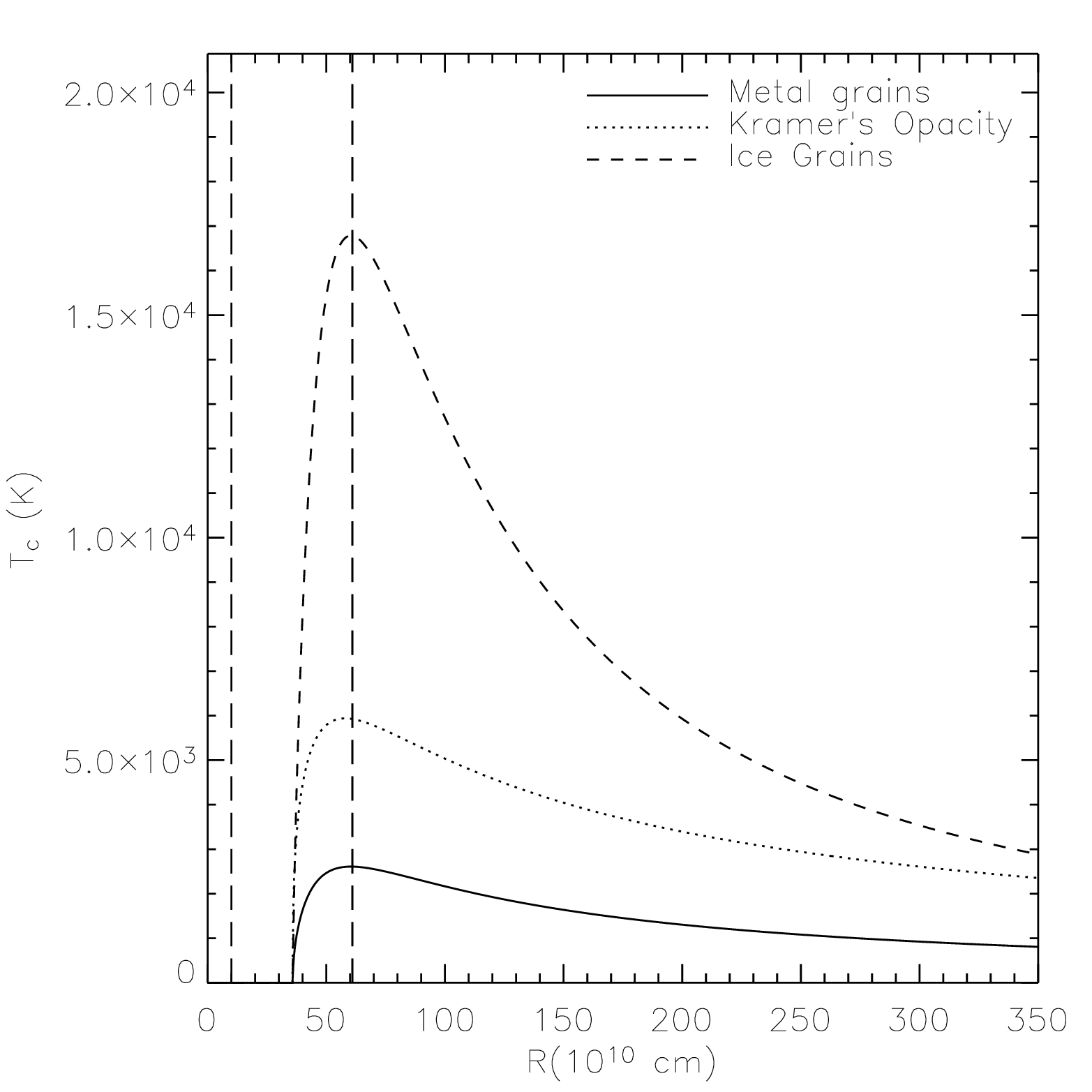}}}
\end{center}
\caption{Analytic model of a typical YSO disc showing surface density
and temperature for three different opacity prescriptions. The disc is plotted from $R=0$ to $R=3.5 \times 10^{12}
\ {\rm{cm}}$. The star has a radius $\rstar=1 \times 10^{11} \
{\rm{cm}}$, mass $\mstar=1 \ \msun$, spin period $P_{\rm{spin}}=3 \
{\rm{d}}$ and a surface magnetic field of $B=500 \ {\rm{G}}$. The disc
has a mass transfer rate of $\mdot=1 \times 10^{-8} \ \msun
{\rm{yr}}^{-1}$, an alpha parameter of $\alpha = 0.01$. In addition to
Kramer's opacity ($\kappa_0 = 5 \times 10^{24}$, $a = 1$, $b = -3.5$), two opacity prescriptions from \citet{bel94} are
illustrated: metal grains ($\kappa_0 = 0.1$, $a = 0$, $b = 0.5$),
and ice grains ($\kappa_0 = 2 \times 10^{-4}$, $a = 0$, $b = 2$). The vertical lines illustrate the stellar radius and the corotation radius.}  
\label{fig:four}
\end{figure*}

\begin{figure*}
\begin{center}
\resizebox{88.1mm}{65.0mm}{
\mbox{
\includegraphics{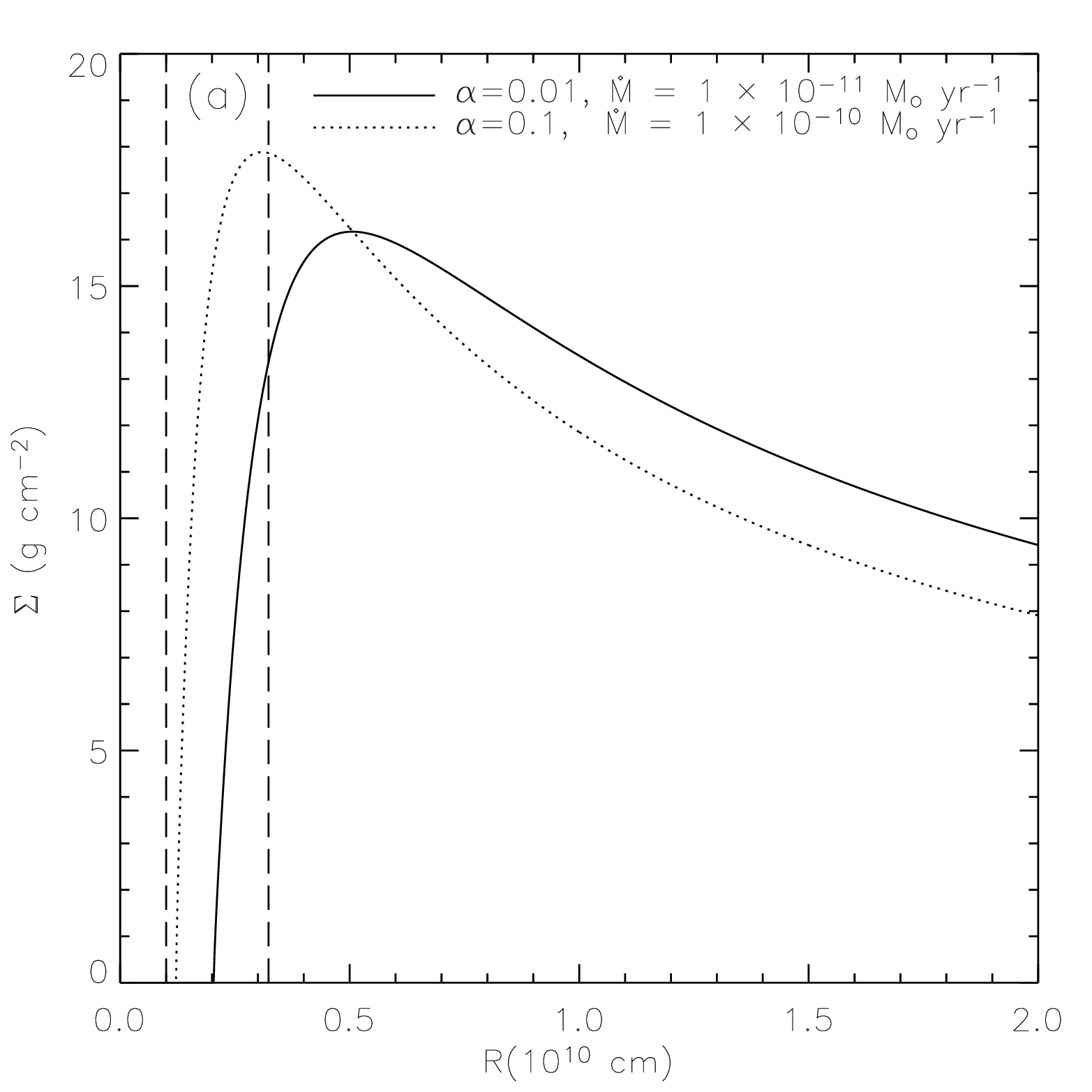}}} 
\resizebox{88.1mm}{65.0mm}{
\mbox{
\includegraphics{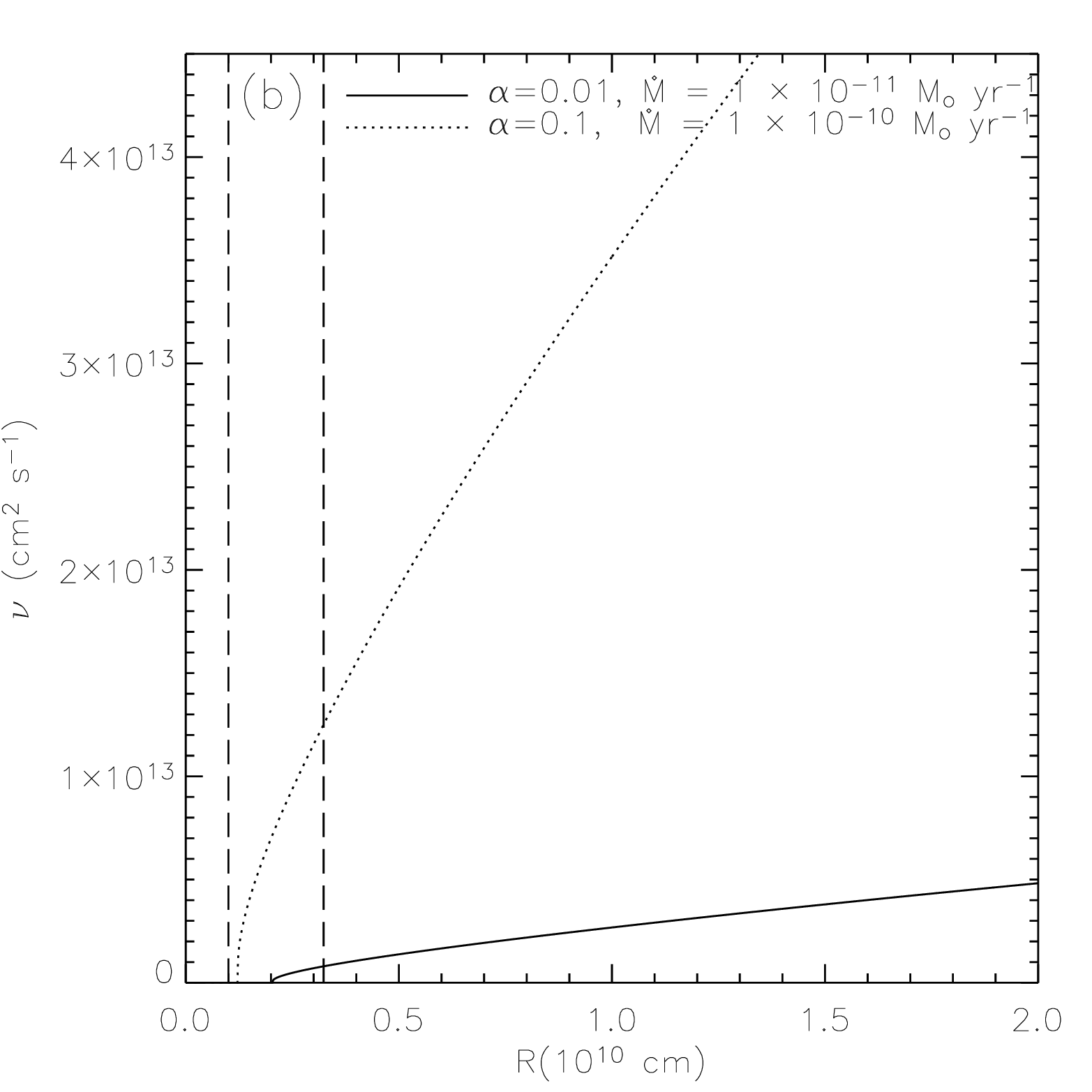}}} 
\end{center}
\caption{Analytic model of a typical dwarf nova disc showing surface density and viscosity for two different values of $\alpha$ and of $\mdot$. The disc is plotted from $R=0$ to $R=2 \times 10^{10} \ {\rm{cm}}$. The star has a radius $\rstar=1 \times 10^{9} \ {\rm{cm}}$, mass $\mstar=1 \ \msun$, spin period $P_{\rm{spin}}=100 \ {\rm{s}}$ and a surface magnetic field of $B=2 \times 10^{3} \ {\rm{G}}$. Kramer's opacity is assumed. The alpha parameter takes values of $\alpha = 0.01$ and $0.1$, representing the disc in quiescence and outburst respectively. The disc has a mass transfer rate of $\mdot=1 \times 10^{-11} \ \msun \ {\rm{yr}}^{-1}$ in quiescence and $1 \times 10^{-10} \ \msun \ {\rm{yr}}^{-1}$ in outburst. The vertical lines illustrate the stellar radius and the corotation radius.}
\label{fig:two}
\label{fig:ob}
\end{figure*}

\begin{figure*}
\begin{center}
\resizebox{88.1mm}{65.0mm}{
\mbox{
\includegraphics{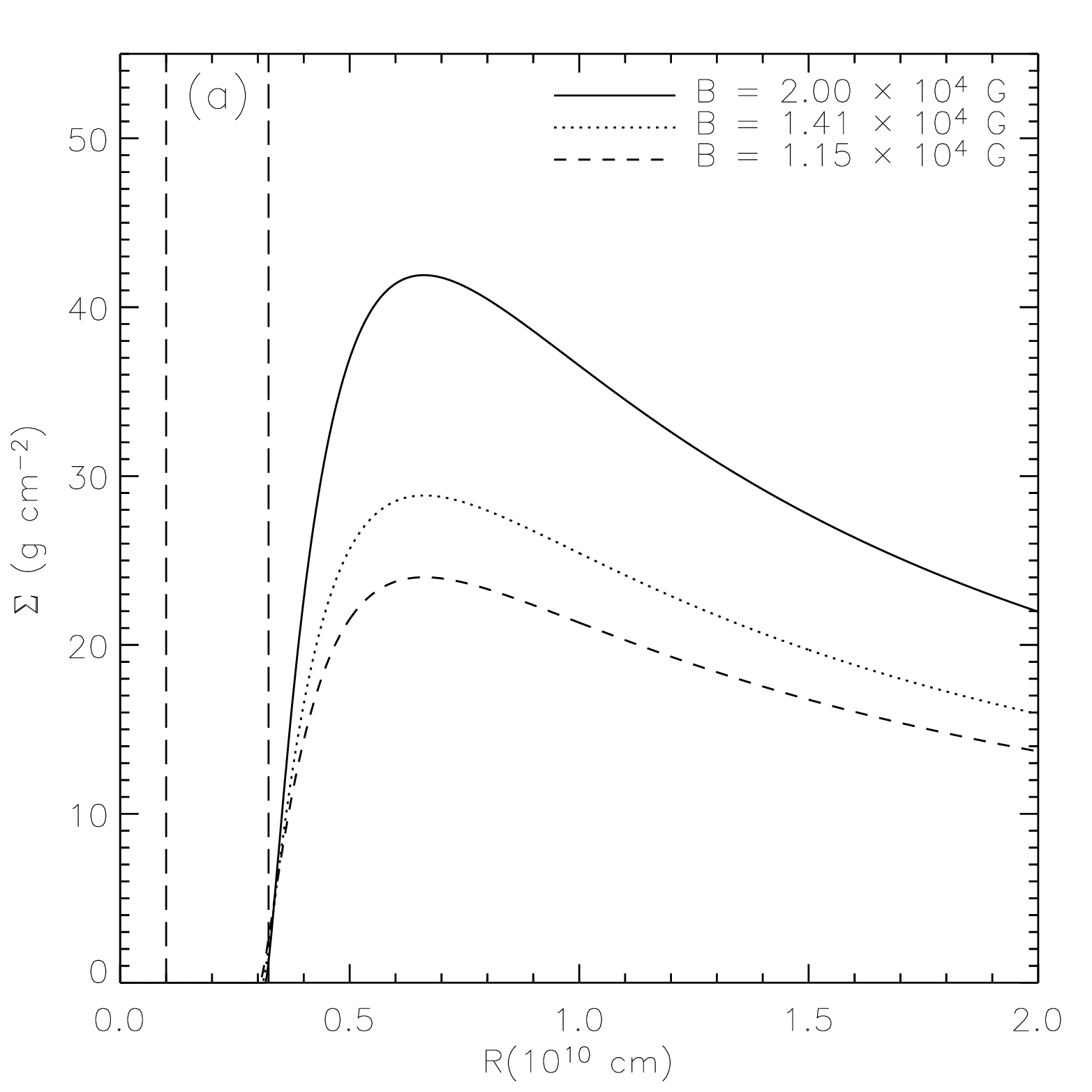}}} 
\resizebox{88.1mm}{65.0mm}{
\mbox{
\includegraphics{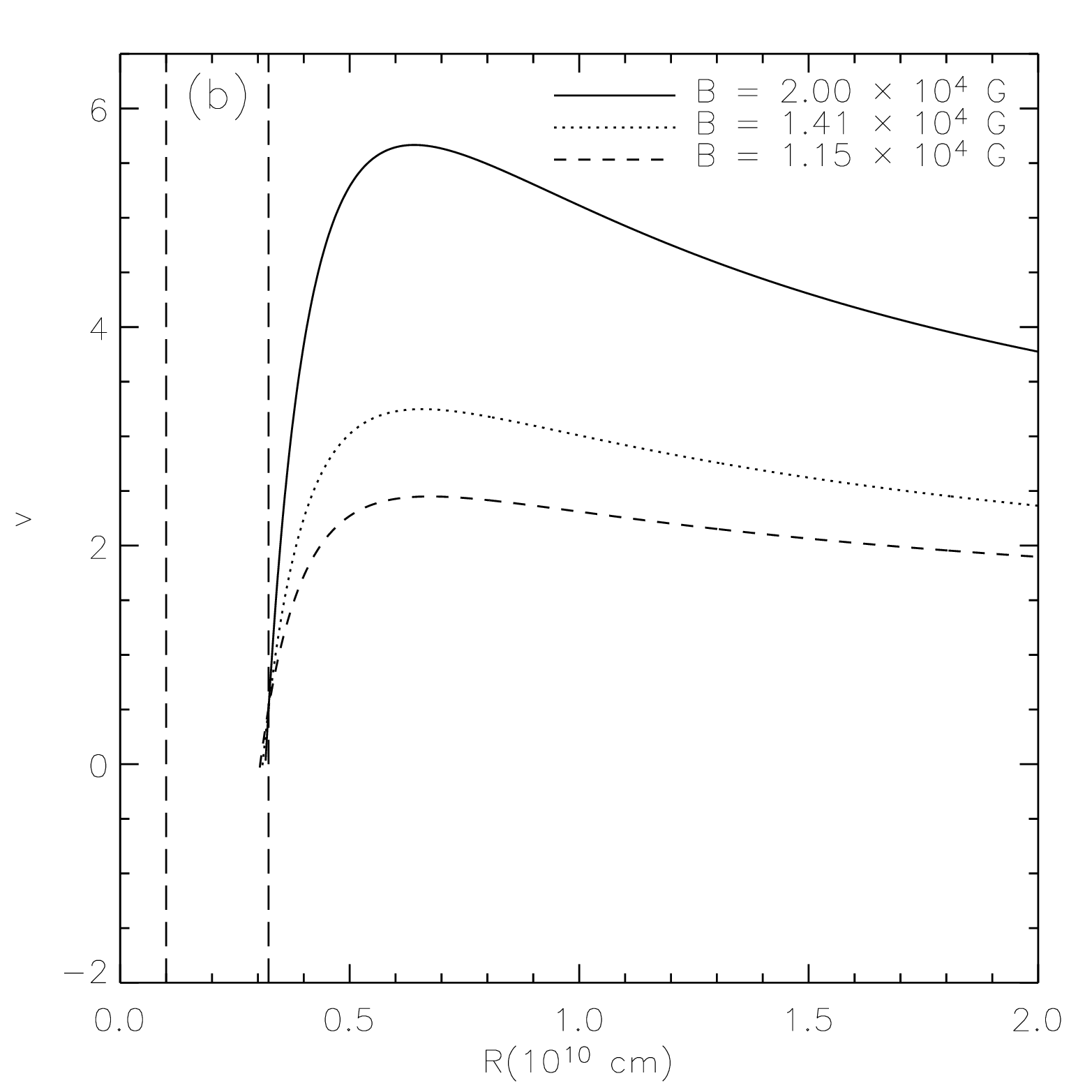}}} 
\resizebox{88.1mm}{65.0mm}{
\mbox{
\includegraphics{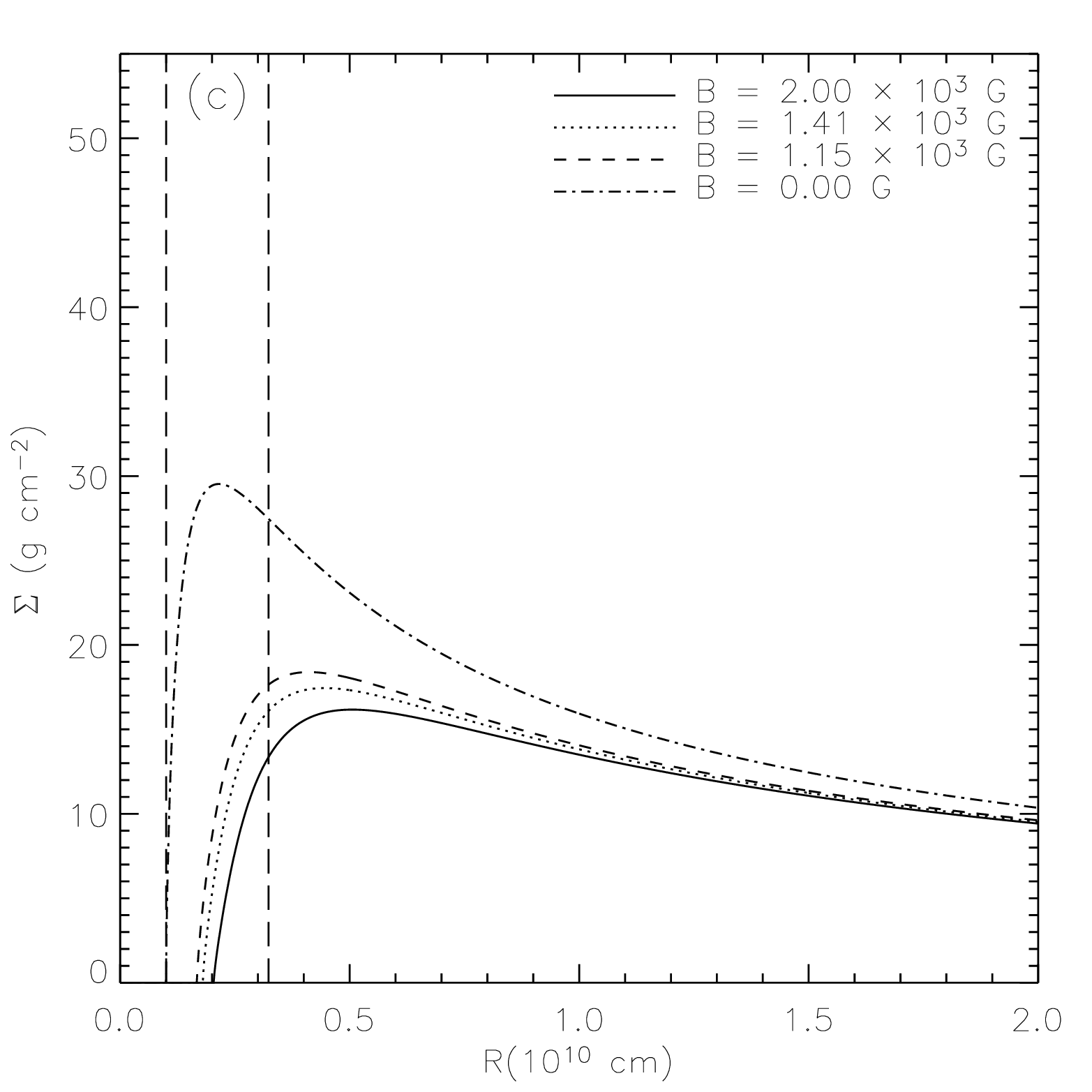}}}
\resizebox{88.1mm}{65.0mm}{
\mbox{
\includegraphics{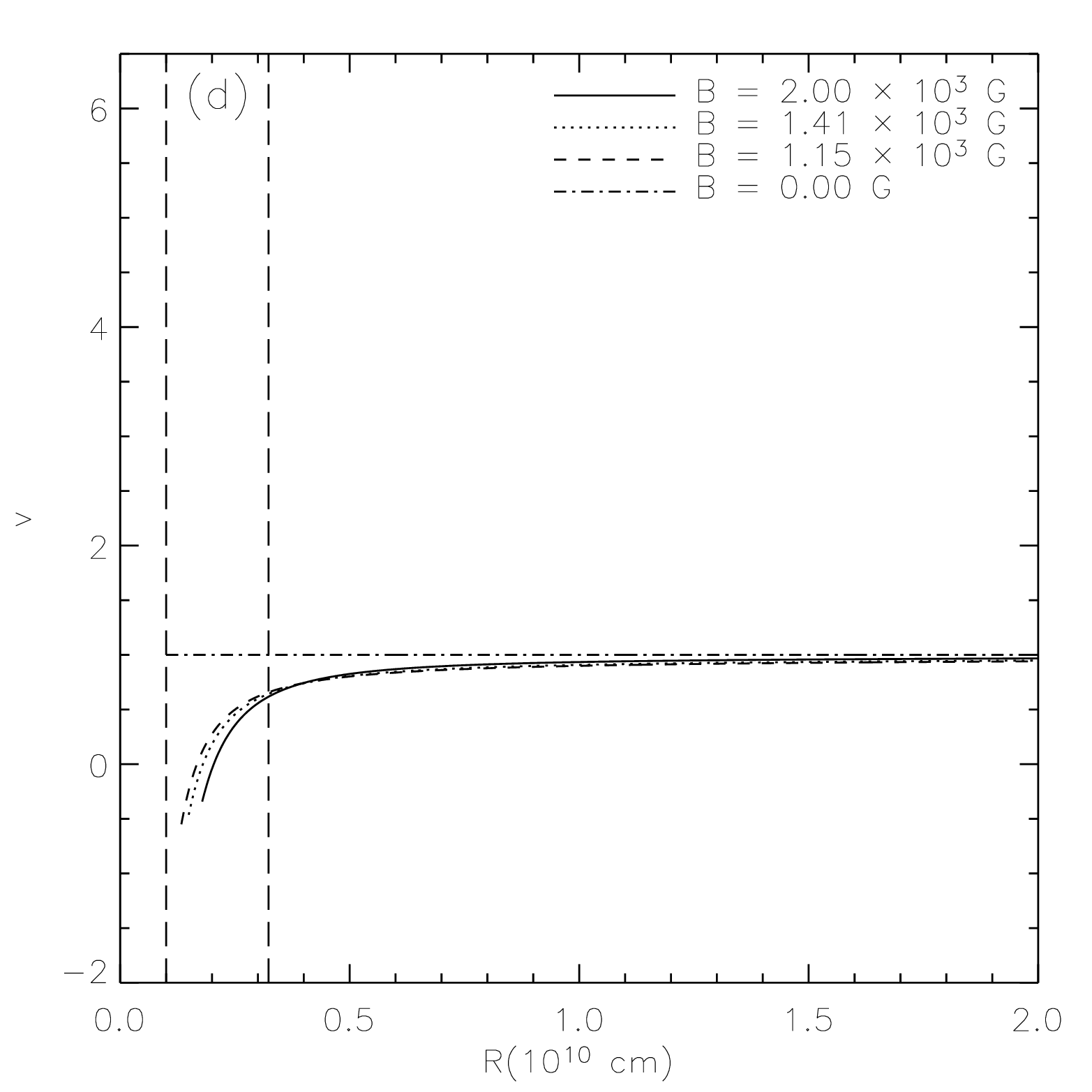}}}
\end{center}
\caption{Analytic model of a typical dwarf nova disc showing surface density and $v$ as functions of radius for a range of magnetic field strengths. The disc is plotted from $R=0$ to $R=2 \times 10^{10} \ {\rm{cm}}$. The star has a radius $\rstar=1 \times 10^{9} \ {\rm{cm}}$, mass $\mstar=1 \ \msun$ and spin period $P_{\rm{spin}}=100 \ {\rm{s}}$. The disc has a mass transfer rate of $\mdot=1 \times 10^{-11} \msun \ {\rm{yr^{-1}}}$ and Kramer's opacity is assumed. The alpha parameter takes the value of $\alpha = 0.01$. The stellar magnetic field ranges from $B=2 \times 10^{4}$ to $1.15 \times 10^{4} \ {\rm{G}}$ in the upper two plots and from $B=2 \times 10^{3}$ to $0 \ {\rm{G}}$ in the lower plots. These values are selected so that the ratio $\beta / \mdot$ is the same as in the three graphs plotted in Fig. \ref{fig:six}, in order that the effects of $\mdot$ and $\beta$ may be seen independently.}
\label{fig:five}
\end{figure*}

\begin{figure*}
\begin{center}
\resizebox{88.1mm}{65.0mm}{
\mbox{
\includegraphics{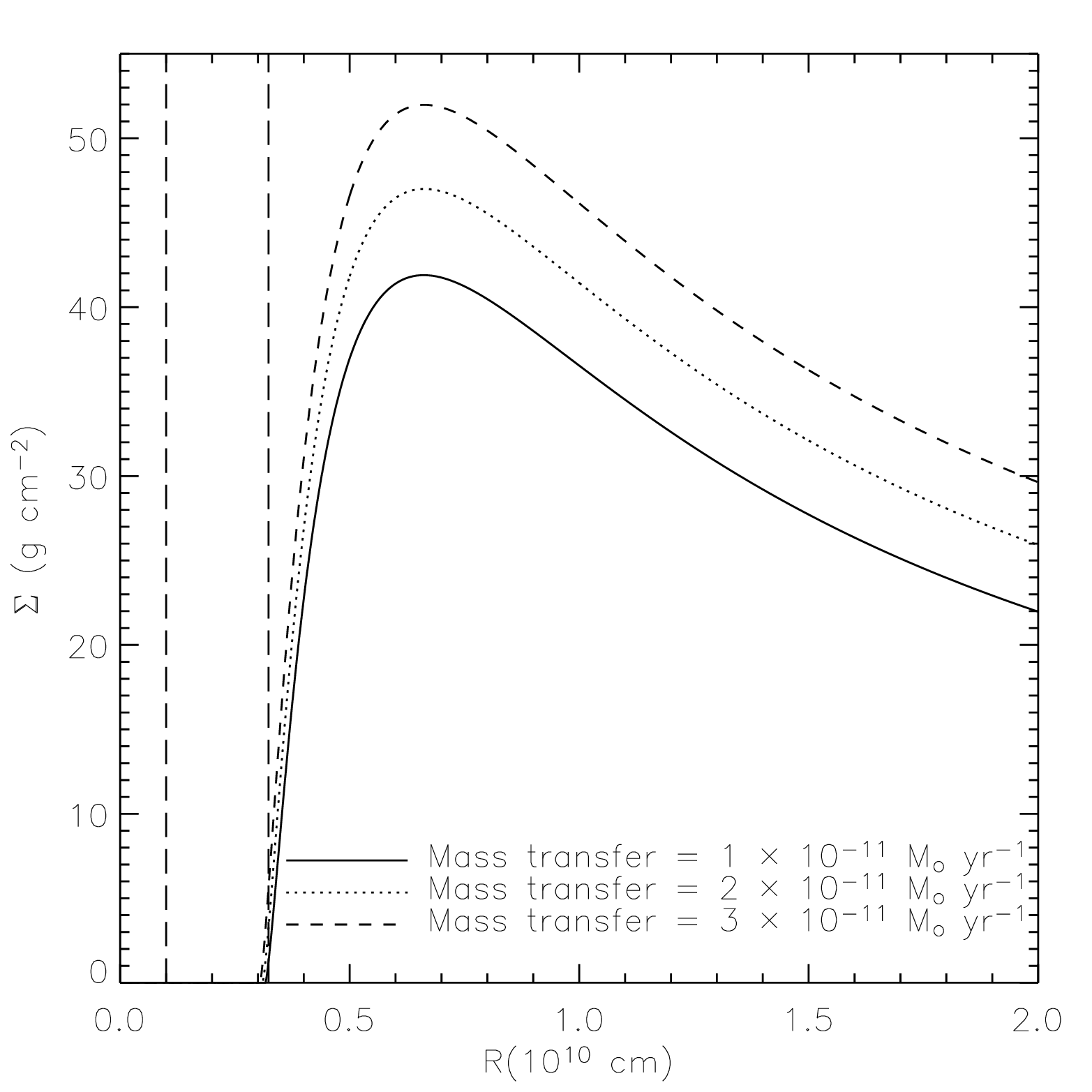}}} 
\resizebox{88.1mm}{65.0mm}{
\mbox{
\includegraphics{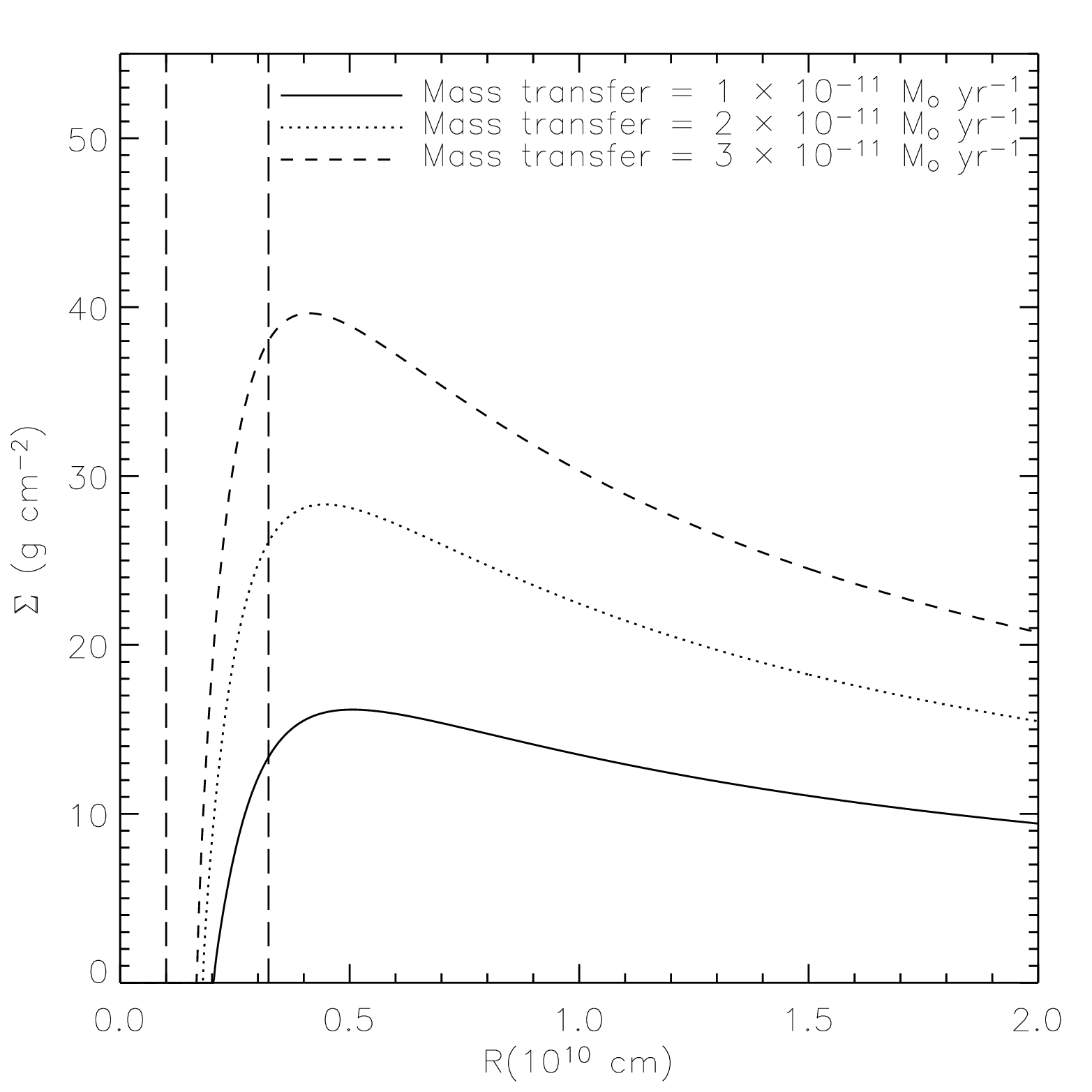}}} 
\end{center}
\caption{Analytic models of a typical dwarf nova disc with a range of mass transfer rates. The disc is plotted from $R=0$ to $R=2 \times 10^{10} \ {\rm{cm}}$.  The star has a radius $\rstar=1 \times 10^{9} \ {\rm{cm}}$, mass $\mstar=1 \ \msun$, spin period $P_{\rm{spin}}=100 \ {\rm{s}}$ and a surface magnetic field of $B=2 \times 10^{4} \ {\rm{G}}$ in the first plot and $B=2 \times 10^{3} \ {\rm{G}}$ in the second. The disc has a range of mass transfer rates from $\mdot=1 \times 10^{-11}$ to $3 \times 10^{-11} \ \msun \ {\rm{yr}}^{-1}$ and Kramer's opacity is assumed. The alpha parameter takes the value of $\alpha = 0.01$. Again, the vertical lines illustrate the stellar radius and the corotation radius.}
\label{fig:six}
\end{figure*}

\begin{figure*}
\begin{center}
\resizebox{88.1mm}{65.0mm}{
\mbox{
\includegraphics{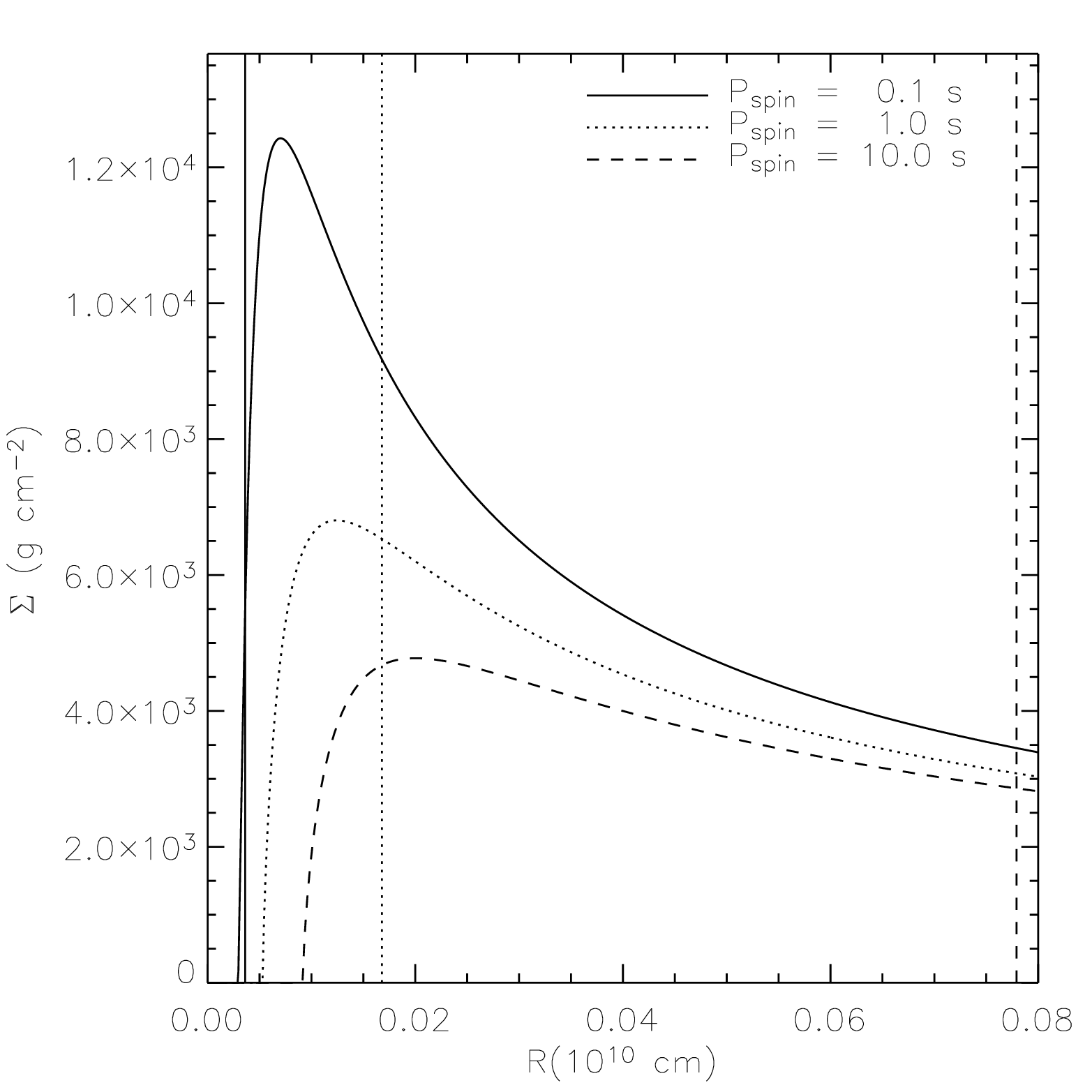}}} 
\resizebox{88.1mm}{65.0mm}{
\mbox{
\includegraphics{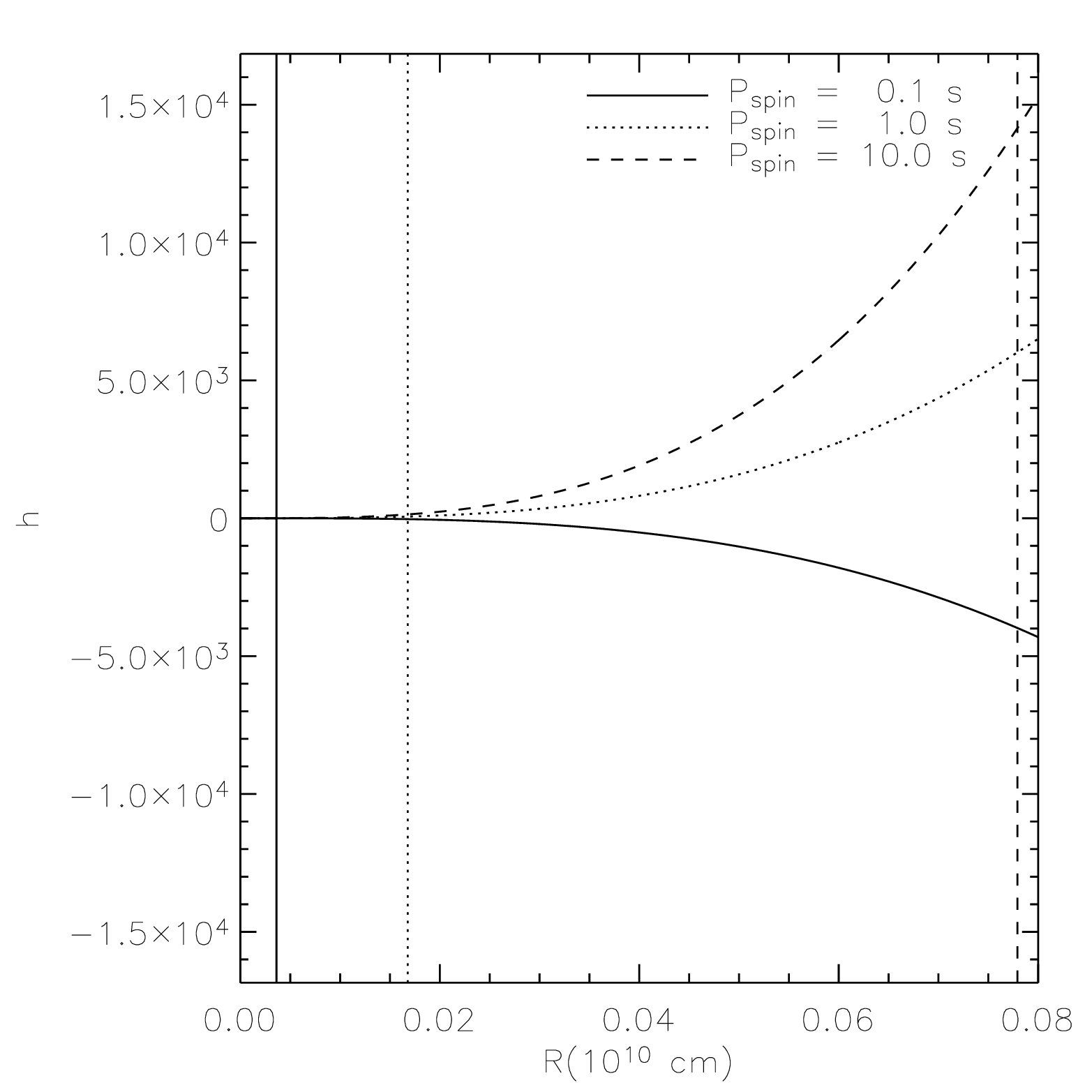}}}
\end{center}
\caption{Analytic model of an X-ray binary disc containing a magnetic neutron star which rotates at different speeds. The disc is plotted from $R=0$ to $R=8 \times 10^{8} \ {\rm{cm}}$, showing surface density and $h$ as functions of radius. The star has a radius $\rstar=1 \times 10^{6} \ {\rm{cm}}$, mass $\mstar=1.4 \ \msun$ and spin periods ranging from  $P_{\rm{spin}}=0.1 {\rm{s}}$ to $10 {\rm{s}}$. The disc has a mass transfer rate of $\mdot=1 \times 10^{-9} \ \msun \ {\rm{yr^{-1}}}$ and Kramer's opacity is assumed. The alpha parameter takes the value of $\alpha = 0.01$. The stellar magnetic field is set to $B=2 \times 10^{8} \ {\rm{G}}$. The vertical lines, which are plotted with the same line style as their associated plots, represent the corotation radii of the primary, at the three different spin periods.}
\label{fig:three}
\end{figure*}

The sample star has typical properties for a low mass YSO so that $\rstar=1 \times 10^{11} \ {\rm{cm}}$, $\mstar=1 \ \msun$, $P_{\rm{spin}}=3 \ {\rm{d}}$. The vertical magnetic field at the surface is $B_{z}=500 \ {\rm{G}}$, which is quite low when compared to recent measurements which give fields of order kG \citep[\eg][]{joh99}. However there is an uncertainty in how $\beta$ relates to field strength, which makes comparison with measured field strengths imprecise. The disc is illustrated at mass transfer rates of between $\mdot=5 \times 10^{-9} \ \msun \ {\rm{yr}}^{-1}$ and $\mdot=2 \times 10^{-8} \ \msun \ {\rm{yr}}^{-1}$. This is a typical $\mdot$ as measured at the star \citep{gul98} and must, in steady state, be the $\mdot$ into every annulus of the disc.  

Fig. \ref{fig:one}a shows surface density as a function of radius. The form is similar to that of the usual non-magnetic Shakura-Sunyaev solution. It differs however in that the inner disc is truncated further out than the stellar surface. In this plot only results at radii greater than $R_{\rm{t}}$ are shown since within $R_{\rm{t}}$ the solution becomes unphysical. In reality this region may contain more complex accretion streams onto the magnetic poles of the star, which are difficult to reproduce in a one-dimensional treatment. Truncation should occur in the region where the magnetic time-scale and viscous time-scale become comparable. In this region the magnetic field is also sufficiently influential to produce effects such as magnetic warping \citep{osu04}. As the accretion rate increases, so does the surface density. As expected, an increased accretion rate also pushes the truncation radii in to smaller radii.

The disc scaleheight (Fig. \ref{fig:one}b) follows an approximately
linear relation with radius, as with the usual Shakura-Sunyaev
solution. However, truncation occurs some distance from the star in
the same manner and at the same radii as with the surface
density. The disc becomes thicker at higher accretion rates, but for realistic parameters remains of the order $H \sim 0.1 R$ so that the thin disc approximation appears to be justified with these parameters. The temperature of the disc mid-plane, as illustrated by
Fig. \ref{fig:one}d, follows a series of curves similar in form to those of the
surface density. This is to be expected since $T_{\rm{c}} \propto
\Sigma^{2}$ for Kramer's opacity. Fig. \ref{fig:one}e plots $|v_{\rm{R}} / v_{\varphi}|$ as a function of radius where $v_{\varphi}$ is assumed to be Keplerian and $v_{\rm{R}}$ is calculated using equation (\ref{eqn:vrmt}). It can be seen that $v_{\rm{R}} \ll v_{\varphi}$ outside the truncation radius, so the assumption that the disc is cool is valid for the region which is modelled. 

The advantage of the solution derived in Section \ref{sec:opa} is that it can be applied to a general opacity case, providing that the opacity law follows the relation given by equation (\ref{eqn:opacitygen}). The discs about young stars are likely to be dusty and cooler than those in most cataclysmic variables or X-ray binaries. The surface density and temperature of such protoplanetary discs are plotted in Fig. \ref{fig:four} for three different opacity laws. The opacity prescriptions for those plots were obtained from \citet{bel94}. The shape of the curves are affected only slightly by the change in opacity, but the magnitude of the densities and of the temperatures are altered by several orders. It is notable, although unsurprising given the form of equation (\ref{eqn:qdef}), that the truncation radius does not vary with opacity. Plotting the temperature of the disc mid-plane provides a useful check of consistency for opacity assumptions. It is clear for example that the inner disc here should not be represented by an opacity law which models the behaviour of ice crystals, since $T_{\rm{c}} > 1 \times 10^{4} \ {\rm{K}}$ in much of the inner disc if such an assumption is made.

In addition to the above-mentioned uncertainties regarding the disc viscosity, the young stellar objects have limitations as test cases for the new disc solution. The study of outburst behaviour is particularly difficult with YSOs. Since only the very inner part of the disc is modelled, the mass transfer rate into the simulated region would in reality be related to the viscous state of the outer disc so that $\alpha$ and $\mdot$ are not independent. This complication is avoided for dwarf novae, where the whole disc can be simulated. In this case the mass transfer rate into the accretion disc is unrelated to the viscous state of the disc. Moreover, the behaviour of dwarf nova discs is in general better understood than that of those around YSOs \citep{war95}. In the hot viscous state dwarf nova discs are very close to the steady state. 

In the case of binaries however, the disc occupies a more complicated potential and an axisymetric model such as the one described in Section \ref{sec:str} must, unavoidably, be an incomplete one. Effects such as tidal resonances cannot easily be modelled in one dimension. A one-dimensional model should however be a reasonable approximation whilst magnetic and viscous forces dominate over tidal ones as they do in the inner disc. 

Fig. \ref{fig:two} shows the steady-state density profiles and
viscosity profiles of a typical dwarf nova disc in a hot, high
alpha; and in a cold, low alpha viscous state. The disc is plotted from
$R=0$ to $2 \times 10^{10} \
{\rm{cm}}$ while the central white dwarf has a radius of $\rstar = 1
\times 10^{9} \ {\rm{cm}}$. The white dwarf is spinning with a period
of $P_{\rm{spin}} = 100 \ {\rm{s}}$ and the surface magnetic field is
$B=2 \times 10^{3} \ {\rm{G}}$, which would normally be insufficient
for the binary to be considered magnetic, since intermediate polar
binaries have fields of $B \sim {\rm{MG}}$ \citep{war95}. It should be noted 
however that magnetic field strengths for white dwarfs with weaker fields 
are not generally well known. The disc opacity is modelled using 
Kramer's law throughout this and subsequent disc models.  

It is interesting, and possibly counterintuitive, that a change in $\alpha$ does not, in itself, cause $R_{\rm{t}}$ to migrate. That this must be the case is very clear however from equation (\ref{eqn:newtrunc1}). In fact, for a given steady-state solution, the only parameters which set the position of the truncation
radius are $\mdot$, $\beta$, $\rstar$ and $P_{\rm{spin}}$. However, the truncation radius is not truly independent of $\alpha$ since, during the change of the viscous state, $\mdot$ also changes and $R_{\rm{t}}$ is a function of $\mdot$.

Although in reality the disc of a dwarf nova does not occupy a true steady state during either outburst or quiescence, it is a reasonable approximation to represent these two phases by two steady states. The outburst cycle can then be interpreted as a cycle between these two steady states. The first plot in Fig. \ref{fig:two} illustrates the density profile
of the disc in the hot and cold states where $\alpha = 0.1$ and $0.01$
respectively while $\mdot = 1 \times 10^{-11}$ and $1 \times 10^{-10}$. The second plot shows disc viscosity for the same two steady states. If the viscous time-scale is given by
\begin{equation}
\label{eqn:visct}
t_{\rm{visc}} = \frac{R^{2}}{\nu}
\end{equation}
then in the hot state the viscous time-scale is reduced from
$t_{\rm{visc}} \sim 1000 \ {\rm{d}}$ to $ \sim 100 \ {\rm{d}}$  at $R = 2 \times
10^{10} \ {\rm{cm}}$. This allows viscous processes to remove the inner
disc material more rapidly and reduce the build up of mass.

It was noted in Section \ref{sec:synthesis} that the new magnetic solution modifies the form of the standard Shakura-Sunyaev model only by the function $v$ which in turn varies from unity according to the ratio $\beta / \mdot$, and by replacing $R_{\star}$ by $R_{\rm{t}}$. This means that when $\beta$ tends to zero, and, $R_{\rm{t}}$ tends to $R_{\star}$, we recover the Shakura-Sunyaev solution and that for high $\mdot$ the magnetic effect becomes less important. However, since $\mdot$ also appears elsewhere in the solution, the effects of varying $\mdot$ and $\beta$ are best examined separately. Figs. \ref{fig:five} and \ref{fig:six} show a cold dwarf nova disc, where the white dwarf has the same properties as in Fig. \ref{fig:two}. In Fig. \ref{fig:five} the effect on surface density of a change in $B$ is illustrated. In \ref{fig:five}a and \ref{fig:five}b $B$ is varied from $2 \times 10^{3} \ {\rm{G}}$ to $1.15 \times 10^{3} \ {\rm{G}}$. The function $v$ is plotted for the same parameters in Fig. \ref{fig:five}b. As has been mentioned, there is no steady state for a magnetic propeller. However the regime illustrated here is that of a `near-propeller'. The truncation radius is very close to its maximum radius, that of the corotation radius, and moves very little with increased field strength. Therefore the great majority of the disc mass lies outside corotation and is propelled outwards. The stronger the field, the greater this tendency will be. If the magnetic advection term is to be balanced by a viscous diffusion term then, for an increasing magnetic field, an increased surface density is required. In this case the function $v$ acts to enhance the surface density over most of the disc. Mathematically, there is no limit to the surface density enhancements which may be obtained in this manner. Physically however, there are two reasons why this is not so. Either the disc will, at some point, reach so high a surface density that a thermal-viscous outburst will occur, or the disc will become very thick and the thin disc treatment will cease to be valid.

For lower field strengths the disc behaves differently. This `strong accretor' regime is illustrated in Figs. \ref{fig:five}c and \ref{fig:five}d. Where $\beta = 0$ we have the Shakura-Sunyaev solution and $v$ retains a value of unity at all radii. As $\beta$ is increased the truncation radius migrates outwards and in this case the surface density decreases throughout the disc. The effect is however more pronounced in the inner disc and the solution tends to the non-magnetic case with increasing $R$, as can clearly be seen from the form of $v$, which never exceeds unity, and so always reduces $\Sigma$.

Fig. \ref{fig:six} shows how the steady-state solution for a dwarf
nova system, which is otherwise identical to that shown in Fig.
\ref{fig:five}, varies when $\mdot$ is changed and $\beta$ is held
constant. In both plots $\mdot$ is varied from $1 \times 10^{-11} \
{\rm{\msun \ yr^{-1}}}$ to $3 \times 10^{-11} \ {\rm{\msun \
yr^{-1}}}$. Care is taken that the ratios $\beta / \mdot$ are, in the first plot, the same as in Fig. \ref{fig:five}a and, in the second plot, are the same as in Fig. \ref{fig:five}c, although the $B=0$ plot has no analogue here. It can be seen that, while the truncation radii are
identical to those in the case where $B$ is decreased, the surface densities are
higher in the case where $\mdot$ is increased. In fact even in the `strong accretor' case, surface density increases with increasing $\mdot$. This is as
expected from the form of equation (\ref{eqn:arbsig}). It is significant that for $\mdot > 2 \times 10^{-11} \  {\rm{\msun \ yr^{-1}}}$ the truncation radii obtained by magnetic truncation are comparable to those of order $\rt \sim 10^{9} \ {\rm{cm}}$ predicted by \cite{kin97} as a result of irradiation. However at lower mass transfer rates the magnetic $\rt$ will become larger than that due to irradiation.

In order to illustrate the effect of a change in $P_{\rm{spin}}$ upon
the disc, a more rapidly rotating star is advantageous. For this
reason we use a neutron star in a low-mass X-ray binary disc. In Fig.
\ref{fig:three} the disc is plotted from $R=0$ to $R=8 \times 10^{8} \
{\rm{cm}}$. This is only a model of the centre of the disc, since this
is where the solution diverges most from the non-magnetic case. The central
star has typical properties for a neutron star in a low
mass X-ray binary, with radius $\rstar=1 \times 10^{6} \ {\rm{cm}}$,
mass $\mstar=1.4 \ \msun$ and a range of spin periods from
$P_{\rm{spin}}=10 \ {\rm{s}}$ to $0.1 \ {\rm{s}}$. The stellar magnetic field is set to $B=2 \times 10^{8} \
{\rm{G}}$ for all plots. The mass transfer rate has a constant value of $\mdot=1 \times 10^{-9} \ \msun \
{\rm{yr^{-1}}}$. The disc is assumed to be in a cool state and to obey
Kramer's opacity throughout. Since we only model the centre of the disc
it is reasonable to ignore tidal effects. The corotation radii here vary according to $P_{\rm{spin}}$ and are marked as vertical lines with the same line style as their associated plots.

The surface density plot shows that an
increased spin rate causes the magnitude to increase and brings
$R_{\rm{t}}$ closer to the star. To understand what causes this it is
useful to examine the form of $h$, defined in equation
({\ref{eqn:hdef}), which represents the potential for departure from
the non-magnetic model as a function of $R$ and of $\rco$. This
function is independent of the $\beta / \mdot$ ratio, which is in any
case held constant throughout Fig. \ref{fig:three}. The function is
made up of two terms, the first of which depends upon the stellar spin
and the second of which does not. The two terms have opposing signs
for $\gamma > 2$. The expression can be interpreted so that the second
term represents the residual accretion effect of a notional
non-rotating magnetic star. For long spin periods, and hence for large
values of $\rco$, it is clear that $h$ tends to this non-rotational
case. The first term in $h$ is
the propeller-like term which becomes more important when $R >
\rco$. The transition between accretor and propeller is not however so
abrupt or straightforward as might be imagined because of the
different powers of $R$ in the two terms. The first case in Fig.
\ref{fig:three} represents a `near-propeller' solution, which has the signature of a negative $h$, which increases surface density. In this case a `near-propeller' solution requires a stronger field than a `strong accretor' solution to truncate the disc at the same radius. This may seem 
counterintuitive, but it agrees with the results of \citet{liv92} and 
can be understood by comparing the mechanisms of 
the magnetic propeller and magnetic accretor. With a given spin period, and disc mass, the magnetic propeller can 
potentially clear a larger hole in the disc, because it is not limited by the
corotation radius. However, because a magnetic propeller has to act in
opposition to viscous processes, and because it causes a build up of mass at 
the edge of the hole, a stronger field is required to maintain the same size 
of hole in 
the propeller case. For faster spinning stars, the corotation radius 
is smaller and so the propeller mechanism operates from closer to the star. 
This causes the truncation radius to become smaller.

\section{Discussion}
\label{sec:discussion}
A new self consistent analytic solution has been developed for a thin accretion disc, under the influence of a central magnetic field. Since this model represents a steady-state disc it is not applicable to all accretion discs, and cannot be used for the magnetic propeller. The model was obtained in a similar manner to the non-magnetic Shakura-Sunyaev model, and indeed the result tends to that non-magnetic case when the magnetic parameter $\beta$ tends to zero. The two models also converge at large radii. The analytic model has also been expanded to include the effects of a range of opacities in addition to Kramer's law. The solution has been applied in the cases of young stellar objects, cataclysmic variables and X-ray binaries.

The new model produces a solution in which the inner disc is frequently truncated further out than the stellar boundary layer. This is caused by magnetically induced accretion. The results depend upon how the position of the truncation radius $R_{\rm{t}}$ varies as a function of the parameters of disc and star. It has been shown that the truncation radius is independent of the opacity prescription used. The magnetic interaction model is uncertain, so that neither the exact values of the magnetic parameters $\beta$ and $\gamma$ nor their dependence on the field strength of the star are so far well known. Further work is therefore required to better constrain the relationship between the surface field and the truncation radius. In addition, a comparison of the position of the truncation radius between the one dimensional solution and three dimensional techniques, such as smoothed particle hydrodynamics, could be used to ensure that the one dimensional inner boundary condition acts in the same manner as the three dimensional analogue.

The critical radius at which thermal viscous outbursts begin varies
almost linearly with radius \citep*{can88}. Therefore a truncated
disc can store more mass before it goes into outburst. As a result
magnetic fields can exert a strong influence on outburst
cycles in accretion discs by forcing outbursts to begin further out in
the disc. This is discussed in more detail by \citet{mat03}. It would
be instructive to perform full outburst simulations in which disc
annuli are permitted to switch between the hot and cold state
according to local triggers, and on a thermal time-scale. Similar work
has been carried out to model dwarf nova outbursts in smoothed
particle hydrodynamics \citep*[\eg][]{tru04}. The steady-state
viscosity prescription should not be applied to a time evolution model
because it becomes inaccurate at low densities. In addition a negative
value of $\nu$, which can occur for $R < R_{\rm{t}}$, will cause equation (\ref{sigmaevol}) to have sinusoidal
solutions. More realistic simulations could be performed by relating
viscosity to surface density or by using a more physically motivated 
viscosity such as that generated by the magneto-rotational instability 
\citep[\eg][]{bal91}. 

Additional future work will involve the generalisation of the new solution to encompass torques from other sources such as planets embedded in the disc. This would have important implications for the study of planet formation and migration in discs. The effect of the magnetic torque on the central star's spin could also be investigated in detail. The long term effect of such spin evolution may be found to have an significant effect on the structure and behaviour of the disc.

\section*{ACKNOWLEDGEMENTS}
\label{sec:ack}
We thank an anonymous referee for very useful suggestions which have significantly improved the paper. Research in theoretical astrophysics at the University of Leicester is
supported by a PPARC rolling grant. OMM gratefully acknowledges
support through a PPARC research studentship. MRT acknowledges a PPARC
postdoctoral fellowship. RS wishes to thank the UK Astrophysical
Fluid Facility (UKAFF) in Leicester for their kind hospitality during a
visit, where parts of this work were initiated, and he would like
to acknowledge the funding of this visit by the EU FP5 programme.
We thank Klaus Schenker and Yohann Grondin for useful discussions.


\end{document}